\documentclass[twocolumn,showpacs,preprintnumbers,amsmath,amssymb]{revtex4}
\usepackage{graphicx}
\usepackage{dcolumn}
\usepackage{bm}
\usepackage{here}
\usepackage{color}

\begin{document}


\title{Time-reversal Characteristics of Quantum Normal Diffusion}

\author{Hiroaki S. Yamada}
\email{hyamada[at]uranus.dti.ne.jp}
\affiliation{Yamada Physics Research Laboratory, Aoyama 5-7-14-205, Niigata 950-2002, Japan}
\author{Kensuke S. Ikeda}
\email{ahoo[at]ike-dyn.ritsumei.ac.jp}
\affiliation{Department of Physics, Ritsumeikan University
Noji-higashi 1-1-1, Kusatsu 525, Japan}

\date{\today}
\begin{abstract}
This paper concerns with the time-reversal characteristics 
of intrinsic normal diffusion in quantum systems.
Time-reversible properties are quantified by the time-reversal
test; the system evolved in the forward direction for a certain period 
is time-reversed for the same period after applying a small 
perturbation at the reversal time, and the separation between the time-reversed 
perturbed and unperturbed states is measured as a function of 
perturbation strength, which characterizes  sensitivity of 
the time reversed system to the perturbation and is called the time-reversal 
characteristic. 
 Time-reversal characteristics are investigated 
for various quantum systems, namely, classically 
chaotic quantum  systems and disordered systems including 
various stochastic diffusion systems. When the 
system is normally diffusive, there exists a fundamental 
quantum unit of perturbation, and all the models exhibit 
a universal scaling behavior in the time-reversal dynamics  
as well as in the time-reversal characteristics, which
leads us to a basic understanding on the nature of 
quantum irreversibility.
\end{abstract}

\pacs{05.30.-d,05.45.Mt,03.65.-w}

\maketitle


\def\tr#1{\mathord{\mathopen{{\vphantom{#1}}^t}#1}} 

\def\ni{\noindent}
\def\nn{\nonumber}
\def\bH{\begin{Huge}}
\def\eH{\end{Huge}}
\def\bL{\begin{Large}}
\def\eL{\end{Large}}
\def\bl{\begin{large}}
\def\el{\end{large}}
\def\beq{\begin{eqnarray}}
\def\eeq{\end{eqnarray}}

\def\eps{\epsilon}
\def\th{\theta}
\def\del{\delta}
\def\omg{\omega}

\def\e{{\rm e}}
\def\exp{{\rm exp}}
\def\arg{{\rm arg}}
\def\Im{{\rm Im}}
\def\Re{{\rm Re}}

\def\sup{\supset}
\def\sub{\subset}
\def\a{\cap}
\def\u{\cup}
\def\bks{\backslash}

\def\ovl{\overline}
\def\unl{\underline}

\def\rar{\rightarrow}
\def\Rar{\Rightarrow}
\def\lar{\leftarrow}
\def\Lar{\Leftarrow}
\def\bar{\leftrightarrow}
\def\Bar{\Leftrightarrow}

\def\pr{\partial}

\def\Bstar{\bL $\star$ \eL}

\def\etath{\eta_{th}}
\def\irrev{{\mathcal R}}
\def\e{{\rm e}}
\def\noise{n}
\def\hatp{\hat{p}}
\def\hatq{\hat{q}}
\def\hatU{\hat{U}}

\def\iset{\mathcal{I}}
\def\fset{\mathcal{F}}
\def\pr{\partial}
\def\traj{\ell}
\def\eps{\epsilon}

\def\noise{n}
\def\hatP{\hat{P}}

\section{Introduction}
The origin of irreversibility in real world should not be attributed only
to the huge amount of the degrees of freedom, and we can expect that 
it can be caused by the complex dynamical structure 
inside the system \cite{prigogine81,ikeda93}.
Indeed, in homogeneously extended quantum systems defined in 
an infinite-dimensional Hilbert space, a stationary irreversible dynamics 
such as stationary diffusion and a uni-directional stationary 
energy flow can be self-organized in the quantum systems even if
the number of degrees of freedom is small \cite{ikeda93,yamada02}.

However, so far, the nature of wavefunctions describing 
 dissipative states has not been well understood, although 
it must have very complex structure.
The purpose of the present work is to explore 
the nature of the quantum state related to the irreversibility 
by using simple quantum systems with a few degrees of freedom. 
To be more specific, quantum irreversibility of diffusive states 
in some quantum systems is investigated.

The normal diffusion realized in homogeneously extended 
quantum systems provides a typical irreversible phenomenon 
in quantum system \cite{yamada99}. Here, by the term normal diffusion we 
mean that the spread of wavefunction which is represented by
the second-order moment $M(t)$ of the coordinate increases
in proportion to time as, $M(t) \propto t$.
A typical example is the quantum Brownian motion modeled 
by a quantum particle coupled 
with the classical heat bath \cite{caldeira83,dittrich90}. 
The heat bath consists of an infinite number of degrees of 
freedom moving on a rapid time-scale, which generates an 
uncontrollable 
microscopic random force. However, the origin of diffusion 
is not always attributed to a random source coming from an 
infinite number of degrees of freedom forming the heat 
reservoir. Indeed, the normal diffusion can be {\it self-induced} 
in the quantum systems with only a few degrees of freedom 
such as coupled kicked rotor \cite{adachi88} and three-dimensional 
disordered systems without any hidden degrees of freedom
\cite{dunlap89,dunlap90,huang01,kawarabayashi96,izrailev00,yamada99}. 

The purpose of our paper is to explore the time-reversibility of
quantum system exhibiting normal diffusion. To quantify
the time-reversibility of the dynamics we measure 
the effect of the external perturbation on the time-reversed 
evolution. It also represents the sensitivity of the time-reversed 
evolution process to the external perturbation.
 There are various types of spreading of wavepacket 
in quantum dynamics. 
One extreme limit is the localized state with $M(t)\propto t^0$,
and the other limit is ballistic motion with $M(t)\propto t^2$. 
The normal diffusion characterized by $M(t)\propto t$ 
is just on the boundary of two extreme types of dynamics.
When we consider stationarity of the quantum dynamics 
the localized states and ballistic motions are not time-stationary process.
On the other hand, the quantum normal diffusion corresponds to one of 
the quantum stationary states. 

In our preliminary report \cite{yamada10}, we demonstrated the 
 result concerning with quantum normal diffusion. 
In the present paper we give more details of the results 
together with discussions for it.
To examine the difference of the time-reversibility 
among the different types of quantum dynamics is 
an another purpose of the present paper.

Unfortunately, we do not have so many quantum systems of 
small number degrees of freedom 
which genuinely exhibit normal diffusion without the help
of any classical random noise source. However, the number of 
quantum systems exhibiting normal diffusion increases if
we include the stochastic quantum systems, namely, quantum 
systems driven by the classical stochastic 
force, into the object of our investigation.
  We will also investigate the time-reversibility 
of such quantum systems exhibiting {\it noise-induced} 
normal diffusion \cite{haken73,capek85}.

Fidelity is frequently used to measure the sensitivity of the quantum dynamics
to the perturbation 
\cite{peres84,benenti02,benenti03,hiller04,mintert05,haug05,gorin06,jacquod09}. 
However, the fidelity is not a
good measure to characterize the complexity and irreversibility of 
quantum motion linking to classical integrability because the fidelity
has no counterparts in classical dynamics \cite{fidelity-classical}.

In the present paper, instead of fidelity, 
 we propose the following {\it time-reversal test} 
to measure the time-irreversibility of quantum dynamics.
First, we evolve an initial quantum state $|\Psi_0>$ forward in time 
by operating the unitary operator $U^T$, where $T$ denotes 
the time step of the evolution, which we hereafter call reversal time.  
  Next, we perturb the evolved 
state by $P$, and evolve the perturbed state backward in time 
by operating the time-reversed unitary operator $U^{-T}$. 
At $t=2T$ the difference between the perturbed time-reversed 
state $|\Psi_0'>=U^{-T}PU^{T}|\Psi_0>$ and the initial 
state $|\Psi_0>$ is measured by the following normalized deviation 
as a function of the perturbation strength $\eta(>0)$,
\begin{eqnarray}
\label{irrev}
 \delta Q = \frac{[<\Psi_0'|\hat{Q}|\Psi_0'>-
<\Psi_0|\hat{Q}|\Psi_0>]}{<\Psi_0|\hat{Q}|\Psi_0>},
\end{eqnarray}
where $\hat{Q}$ is an appropriate observable related to the coordinate in
which the diffusion takes place. The quantity $\delta Q$ observed as a 
function of the strength $\eta$ of the perturbation $P$ is called
the {\it time-reversal characteristic}. 

The above time-reversal test is immediately related to 
the fidelity of the quantum systems frequently 
estimated in Loschmidt echo experiments, which measures the 
decoherence\cite{peres84,benenti03,mintert05} 
by the distance between the quantum states
with perturbed and unperturbed time-reversed 
dynamics \cite{fidelity}.

The above time-reversal characteristics were introduced by 
one of the authors in 1995 as a measure quantifying the 
sensitivity of quantum dynamics to perturbation \cite{ikeda96}.
This has been motivated by a remarkable stability of quantum 
chaos systems compared with its classical counterpart, which
was first demonstrated by Soviet-Italy group \cite{chirikov88,shiokawa95}. 
A quantum measurement-theoretic discussion about the significance of 
the time-reversal tests is given in Sect.II and appendix C.

The main results in the present paper are summarized as the 
following remarkable universal features of time reversal characteristic.

{\it Result 1:} 
In the time reversal characteristic of quantum normal diffusion
there always exists a universal characteristic strength of $P$ called 
the least quantum perturbation unit (LQPU), which is proportional to Planck
constant \cite{ikeda96}.  For the perturbation strength larger than 
the LQPU the normal diffusion is irreversible in the sense that 
it completely loses its ability to return to the initial state.

{\it Result 2:} 
In deterministic quantum maps the time reversal characteristic 
of quantum normal diffusion converges to a universal curve
independent of the details of the systems in the large limit 
of the reversal time. Moreover, this universality and the 
stationarity of diffusion process are unified to  
a universal scaling of the time reversed dynamics itself. 
On the other hand, the way of convergence toward the universal
characteristics depends on the details of the systems.

{\it Result 3:} The universal features observed for the
deterministic quantum maps exactly hold also for the normal 
diffusion process of stochastically perturbed quantum maps 
if the strength of the stochastic random force is large 
enough to realize a sufficiently rapid normal diffusion.

We note here that the presence of LQPU was first claimed by one of the
authors and its existence was confirmed  for the time-reversal
characteristics of standard map Ref.\cite{ikeda96}. Later the existence 
of LQPU was verified within the limitation of lowest order 
perturbation theory \cite{sokolov08}.

The above universal features in the quantum region are first discussed
for deterministic quantum systems, but it turns out that they are
also exactly valid for stochastic quantum systems.
It should be remarked that what we mean by the term "deterministic" 
quantum diffusion is the normal diffusion induced by the genuine 
quantum dynamics of the system which does not contain any explicit 
classical random force. We distinguish it from the noise-induced 
diffusion exhibited by quantum system driven by classical random force
such as quantum Brownian motions.
Further it should be noted that we refer to the diffusion as even 
if the normal diffusion $M(t) \propto t$ is maintained only 
within a limited time range. Indeed, some quantum systems such as 
one-dimensional quantum kicked rotor cannot show diffusion dynamics 
over an infinitely long time. However, if the diffusion dynamics has a 
well-defined time range on which a well-behaved $t-$linear  
dependence is maintained, we can use it as an example of quantum normal diffusion.

As mentioned above, we add the noise-driven quantum systems to our menu, 
and compare the time-reversal characteristics with those in the 
deterministic normal diffusion systems. 
It is shown that 
the time-reversal characteristics significantly deviate from 
the universal one when the strength of the random force is not 
large enough.

We further extend our investigation of time-reversal characteristics
to quantum dynamics different from the normal diffusion, such as 
localization, subdiffusion and ballistic motion. 
We discuss a seemingly 
paradoxical result that the localized states are more
sensitive to the perturbation than the normally diffusive states.
On the other hand, the ballistic motion is entirely time-reversible 
in a sense that the time-reversal characteristics approaches to zero 
as the reversal time $T$ increases.

The outline of the present paper is as follows. 
In Sect.\ref{sect:model} model systems examined in the present work 
are introduced. In addition, we demonstrate numerically 
typical examples of quantum normal diffusion and confirm
its Markovian property. Section \ref{sect:irrev} is the core part 
of the present work in which the main results of time-reversal
characteristics summarized as the {\it Result 1} and {\it Result 2} 
are clarified for deterministic quantum systems showing normal 
diffusion. 
In Sect.\ref{sect:noise} the main results mentioned above are
examined for the stochastic quantum systems, which exhibit normal 
diffusion if it is driven by classical random force.
The last section is devoted to summary and discussion.

Appendixes are devoted to providing
a lot of numerical data which supplements the main results
from several points of view.  
Accordingly the busy readers can get the essential claims
of the present work by reading only the main text.
Those who are very interested in the topics of dissipation 
are recommended to consult the appendices.

\section{Model systems exhibiting quantum diffusion} 
\label{sect:model}
\subsection{Model systems}
\label{subsect:model} 
Typical examples of deterministic quantum systems with normal 
diffusion are seen in quantum chaos systems, quantum disordered systems
and so on. In the present paper, we treat quantum systems defined on 
discrete-time, which are often called quantum maps, because it allows a
very long-time evolution with numerical accuracy.
We also deal noise-driven quantum maps, which exhibit normal
diffusion, in comparison with the deterministic quantum maps.
We will discuss the time-continuous quantum systems with 
normal diffusion in a separate paper \cite{yamada11}.

As the quantum map, we use the following form of unitary operator, 
\begin{eqnarray}
\label{unitary}
   \hatU  = \e^{-i\frac{H_0(\hatp)}{2\hbar}}\e^{-i\frac{V(\hatq)}{\hbar}}
\e^{-i\frac{H_0(\hatp)}{2\hbar}}, 
\end{eqnarray}
where $H_0(\hatp)$ and $V(\hatq)$ represent 
translational$-$(or rotational$-$)kinetic energy 
and potential energy, respectively.
Here $\hatp$ and $\hatq$ are momentum and positional operators, respectively.

\subsubsection{Normally Diffusive deterministic quantum maps}
\label{subsubsect:normaldiff}
Standard map (SM) is a typical deterministic quantum map 
which is given by 
\begin{eqnarray}
\label{SM}
   H_0(p)=\frac{\hatp^2}{2}, ~~~~~~~~~V(q)=K \cos \hatq.
\end{eqnarray}
An important feature of SM is that it has the classical
limit for $\hbar\to 0$.
The position space is taken defined as $0\leq q\leq 2\pi$, and 
a periodic boundary condition $\Psi_t(q)=\Psi_t(q+2\pi)$ is
imposed on the wavefunction $\Psi_t(q)=U^t\Psi_0(q)~~(t\geq 0)$. 

In the classical limit $\hbar \rightarrow 0$, its dynamics shows
nearly integral motion on Kolmogorov-Arnold-Moser (KAM) torus 
for the nonlinear parameter 
$K \ll 1$, while it exhibits an unlimited normal diffusion 
for  $K \gg 1$ in the momentum space $-\infty<p<+\infty$, 
where the diffusion constant $D=D_{cl} \sim K^2/2 $ for $K >> 1$.
The quantum counterpart exhibits the classical normal diffusion
within a limited time range $t \ll D/\hbar^2$;
the diffusion is suppressed and localized for $t \gg D/\hbar^2$.
In the present paper, we investigate time-reversal characteristics 
of normal diffusion for $t \ll D/\hbar^2$, which can be made arbitrarily
large by taking $\hbar << 1$ for $K>3$.

We consider a quantum disordered system on one-dimensional discrete lattice
($q \in {\bf Z}$) as the next example. Unlike SM, it has no classical
counterpart. The system is given by 
\begin{eqnarray}
\label{AM}
   H_0(p) &=& \Delta \cos (\frac{\hatp}{\hbar})  \nonumber \\
  &=& \Delta \frac{\e^{\partial/\partial q}+\e^{-\partial/\partial q}}{2}, 
\nonumber \\
  V(q) &= & v_q, 
\end{eqnarray}
where $H_0(p)$ describes hopping between nearest neighbour sites and the 
on-site potential $v_q$ takes random value uniformly distributed over 
the range $[-W, W]$.  
The transfer operator $\cos(p/\hbar)=(\e^{\pr/\pr q}+\e^{-\pr/\pr q})/2$
evidently does not have a classical limit, and so the model has no classical 
counterpart as mentioned above.
The unitary evolution operator given by Eq.(\ref{unitary}) 
reduces to 
\begin{eqnarray}
\hatU_{AM} &=& \e^{-i\Delta(\cos(p/\hbar)+ v_q/\Delta )/\hbar},  
\end{eqnarray}
in the limit $\Delta \to 0$ with $W/\Delta=const.$, 
where the Hamiltonian $H_0+v_q/W$ is nothing more than the one-dimensional Anderson
model. Thus the model given by Eq.(\ref{AM}) is a map version 
of the one-dimensional Anderson model, and so
we refer to it as the Anderson map (AM), and we take the special
choice $\Delta=1$ hereafter  \cite{yamada04}. 

As commonly seen in the one-dimensional Anderson model, the initially localized
wavepacket spreads in the $q$-space but is finally localized as time 
elapses. However, it was shown that the quantum motion 
exhibits a well-defined normal diffusion in $q-$space when we replace 
the static disordered potential  $V(q)$ with the following harmonic 
time-dependent one \cite{yamada04}, 
\begin{eqnarray}
\label{PAM}
   V(q,t)=v_q \{1+  \sum_{i=1}^M \epsilon_i \cos{\omega_i t} \}, 
\end{eqnarray}
where $M$ and $\epsilon_i$ are the number of the frequency component and 
the strength of the perturbation, respectively, and
we call the model with potential given by Eq.(\ref{PAM}) 
the perturbed Anderson map (PAM).

The external harmonic perturbation is equivalent to 
the coupling with quantum linear oscillators, and the model with Eq.(\ref{PAM})
can be transformed into a one-dimensional discrete lattice system coupled with 
$M$ quantum linear oscillators \cite{yamada02}.
In the following numerical calculation, we take 
$W=1.0$ or $0.5$ and $\epsilon_i=\frac{\epsilon}{\surd M}$, for simplicity, 
and take incommensurate numbers $\omega_i \sim O(1)$ as the frequency set.

\subsubsection{Normally diffusive stochastic quantum maps}
\label{subsubsect:stochasticdiff}
Unfortunately, we do not know many deterministic quantum system which
shows rigorous normal diffusion without any stochastic perturbation. 
However, once we turn our attention to quantum systems perturbed by 
a stochastic random force varying at random from step to step,
many one-dimensional quantum systems show normal diffusion.
For example, if the potential in Eqs.(\ref{SM}) and (\ref{AM}) are perturbed by
a stochastic random perturbation, i.e., 
\begin{eqnarray}
\label{SSM}
   V(q, t) = V(q) (1 + \epsilon \noise_t),~~~~~~~~<\noise_t \noise_{t'}>=\delta_{tt'},
\end{eqnarray}
then the corresponding SM and AM exhibit normal diffusion 
irrespective of the magnitude of $\eps$ (if $K$ is large enough).

Moreover, it is well-known that one-dimensional Hamiltonian $H=\cos(p/\hbar)+V(q,t)$ driven by 
spatio- and temporal-uncorrelated noise written as
\begin{eqnarray}
\label{HM}
   V(q, t) = \eps \noise_{q,t},~~~~~<\noise_{q,t}\noise_{q',t'}>=\delta_{qq'}\delta_{tt'}, 
\end{eqnarray} 
shows normal diffusion.
This model has been introduced by Haken and coworkers in the context of exciton
migration in solid, and the diffusive behavior has been well-investigated.
 In addition to the stochastically driven SM \cite{arnold98} and AM, 
we use the quantum map version of the Haken-Strobl model \cite{haken73} as a prototype
with normal diffusion, which is given by, 
\begin{eqnarray}
 \hat{U}=\e^{-i\cos(\hatp/\hbar)/2\hbar}
\e^{-i\noise_{qt}/\hbar}\e^{-i\cos(\hatp/\hbar)/2\hbar}. 
\label{kickedHaken}
\end{eqnarray}
We call the map version the Haken map (HM), hereafter.

\subsubsection{Subdiffusive and localizing quantum maps, superdiffuisive quantum maps}
\label{subsubsect:loc-sub-super}
The quantum maps introduced above also exhibit subdiffusive behavior, 
$M(t)\propto t^{\alpha} (0<\alpha<1)$, 
and localization ($\alpha=0$), depending on the value of the control 
parameters $K$, $\hbar$, and $\epsilon_i$.
Moreover, we can obtain the Bloch map by replacing the random potential with
a periodic one in AM, in which the wavepacket shows a ballistic motion.
We compare the time-reversal characteristics in the cases of
localization, subdiffusion and ballistic motion, with those of 
normal diffusion exhibited by the models in the previous subsection.

\subsection{Typical quantum diffusions} 
\label{subsect:bayse}
\begin{figure}[htbp]
\begin{center}
\includegraphics[width=9cm]{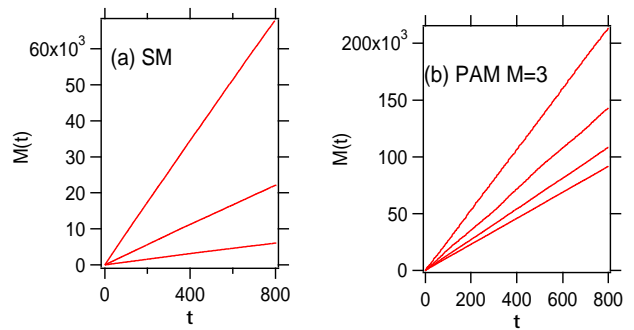}
\caption{\label{fig1-moment} 
(Color online)
Time-dependence of MSD of SM and PAM with normal diffusion. 
(a) SM with $\hbar=\frac{2\pi 243}{2^{18}}, \frac{2\pi 975}{2^{20}}, 
\frac{2\pi 1947}{2^{21}}$ and
 $K=4.2, 6, 12$ from below, respectively.
(b) PAM with $M=3, W=1.0$ and $\epsilon=0.3,0.5,1.0$ from below.
The top curve is a result for $W=0.5, \epsilon=0.5$.
In case of PAM, we usually take the ensemble average over
100-200 different samples of on-site randomness. (No sample
average is taken for SM.) We set $\hbar=1/8$ 
for PAM throughout this paper.
}
\end{center}
\end{figure}

\begin{figure}[htbp]
\begin{center}
\includegraphics[width=9cm]{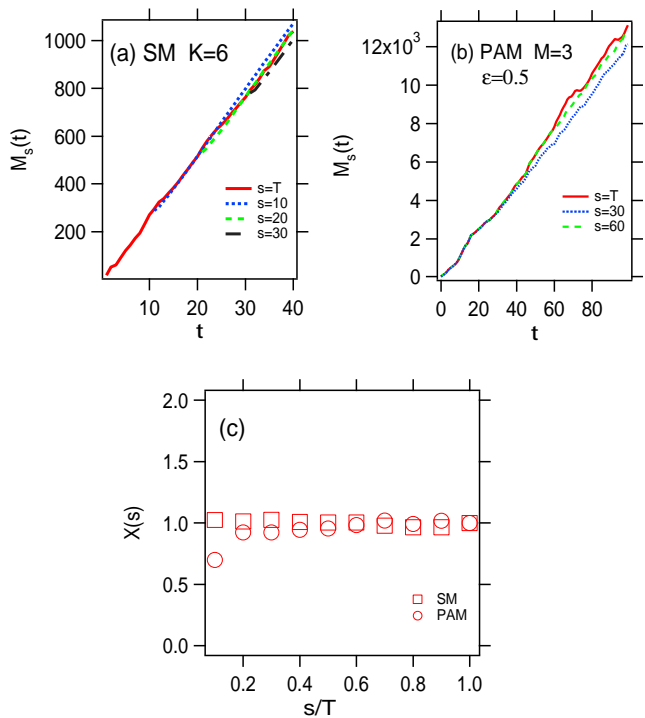}
\caption{\label{fig2}
(Color online)
Time-dependence of MSD with observation at several intermediate
times $s$. $s=10,20,30$ for SM, and $s=30,60$ for PAM.
(a)SM with $K=6$, $\hbar=\frac{2\pi 35}{2^{17}}$.
(b)PAM with $M=3$, $\epsilon=0.5$. 
(c)Index $X(s)=M_s(T)/M(T)$ as a function of $s/T$ for SM and PAM.
The curve for $s=T$ shows time-dependence of MSD without intermediate
observation as a reference. In this case we do not 
take ensemble average for PAM.
}
\end{center}
\end{figure}

Let $x$ be the coordinate of the space in which the diffusion takes
place; namely $x=p$ in SM and $x=q$ in PAM and HM.
We monitor time-dependence of mean square displacement (MSD) of 
wavepacket initially localized on a point at $t=0$, i.e. $<n|\Psi_0>= \delta_{n,0}$,
\begin{eqnarray}
  M(t) = <(\Delta x)^2> = \sum_{x}P(x,t)(x-<x>)^2,  
\end{eqnarray}
where  $<...>$ indicates quantum mechanical average for 
the quantum mechanical probability $P(x,t)=|\psi(x,t)|^2$, 
for example, $<x>=\sum_{x}xP(x,t)$.
In addition, in cases of AM and HM, we also take an ensemble 
average $<...>_{\Omega}$ over different on-site randomness 
and/or stochastic random force although it is not shown 
in equations to avoid complication of the symbols. 
 In Fig.\ref{fig1-moment} we show typical examples of the quantum normal 
diffusion observed for SM and PAM. The diffusion constant $D$ 
increases with the increase of the control parameters; $K$ in SM 
and $\epsilon/W$ in PAM.
%
%
In the realization of quantum normal diffusion, the first basic question 
is whether the square of quantum wave function can be identified
with a classical probability distribution.
First of all, we examine the Bayesian property of the quantum probability
distribution
by investigating whether or not the quantum transition probability 
$P(t,x \rar x')= |<x|U^t|x'>|^2$ at time $t$ satisfies Bayes theorem
\begin{eqnarray}
\label{bayse}
   P_s(t,x \rar x') = \sum_{x''} P(s,x \rar x'')P(t-s,x''\rar x').  
\end{eqnarray}
The well established the Bayesian property implies that the time-evolution 
of the wavepacket of the quantum mappings obeys a stationary Markov process. 
In Fig.\ref{fig2}(a) and (b), instead of time-dependence of probability distribution, 
we compare the time-dependence of MSD computed using the probability
for the actual time evolution, i.e., $M(t)=\sum_{x'}(x'-x)^2 P(t,x\rar x')$,  
and MSD computed using the probability of Eq.(\ref{bayse}), i.e.
\begin{eqnarray}
  M_S(t)=\sum_{x'} P_s(t,x \rar x')(x'-<x>_s)^2 , 
\end{eqnarray}
where $<...>_s$ indicates quantum mechanical average for 
the probability $P_s(t, x \rar x')$.
Figure 2(c) shows an index $X(s)$ as a function of 
the intermediate time $s$ defined as,  
\begin{eqnarray}
    X(s) = \frac{M_s(t)}{M(t)}.
\end{eqnarray}
Figure \ref{fig2}(c) shows that $X(s)$ is equal to unity except
for tiny fluctuation independently of the intermediate time $s$,
which indicates that these pure quantum evolution processes can 
be well approximated by a stationary Markov process. We stress that
in the present test we do not take the ensemble average over 
the random on-site potential. 
%
%
In an ordinary quantum process, Eq.(\ref{bayse}) does not in general 
hold because of quantum coherence.  Even if a measurement of 
$x$ is done at the intermediate time $s$ and so the wavefunction 
shrinks to one of the eigenstates of $x$, the Bayesian property do 
not in general hold for quantum probability. A surprising fact 
is that in the present cases the Bayesian property is valid without 
any measurement.  This means that the conversion of a quantum 
wavefunction into a classical probability occurs {\it spontaneously} 
in the quantum evolution process. 

On the other hand, such stationary Markovian property can not be
observed if localization or subdiffusion takes place instead of
normal diffusion.  The details of the result are given 
in appendix \ref{app:PAM-curve}.

\section{Normally diffusive deterministic quantum maps}
\label{sect:irrev}
This is the main section of this paper.

\subsection{Time-reversal test}
\label{subsect:irrev-test}
As seen in the last section, a normal diffusion obeying a stationary
Markov process is realized in some simple quantum map systems.
The presence of normal diffusion generally suggests that the underlying 
quantum dynamics is complex enough to lose the past memory, and
further this complexity more or less reflects the instability of 
underlying dynamics. 
However, note that the quantum normal diffusion does not always correspond to
instability in the quantum dynamics. Normal diffusion appears  
in the noise-driven (nearly) integrable systems without instability.
We discuss this point in Sect.\ref{sect:noise}.

We quantitatively characterize the instability of
underlying quantum dynamics in terms of the sensitivity of time-reversed 
dynamics to the perturbation applied at a reversal time. 
This method provides a powerful tool for measuring the instability 
of quantum dynamics which have no corresponding classical orbit.

The time-reversal test is executed following the three steps mentioned below.
First the initial state is evolved forward in time by operating the 
evolution operator $U^t$ until a reversal time $t=T$. At the reversal 
time $t=T$, a perturbation $\hat{P}(\eta)$ 
is applied, and next the perturbed state is evolved backward in time  
by operating the time-reversed evolution operator $U^{-T}$.
As introduced in Eq.(1), the relative irreversibility 
$(Q_\eta(2T)-Q_0(2T))/Q_0(T)$ as a function of the perturbation strength
$\eta$ is considered as a measure of the sensitivity of the quantum
state to the perturbation. 
Here $Q_\eta(2T)$ means physical quantity $Q$ when the perturbation 
with the strength $\eta$ is applied at 
the reversal time $T$, and  
$Q_0(2T)$ denotes the value of $Q$ in the unperturbed case, $\eta=0$.
There, however, are various ways to measure the distance between the 
quantum states. We adopt second moment $M(t)=\sum_{x} (x-<x>)^2
|\Psi(x,t)|^2$ of the wavepacket as 
the quantity $Q$, because it captures the most characteristic feature of 
normal diffusion, and then we use the notation $\irrev$ for the
time-reversal characteristics measured by MSD $M(t)$ as,
\begin{eqnarray}
\label{eq:irrev}
   \irrev(\eta)=\frac{|M_\eta(2T)-M_0(2T)|}{M_0(T)}.
\end{eqnarray}
As the perturbation $\hat{P}(\eta)$ applied at $T$, we mainly use the
$\eta-$shift operator, which shifts the wavepacket by $\eta$ 
in the space $y$ canonically conjugate to the diffusion space $x$ 
(namely $y=q$ for $x=p$ as in SM, and $y=p$ for $x=q$ as in PAM and HM),
\begin{eqnarray}
\label{pershift}
   \hat{P_y}(\eta)=\exp \{i\eta \hat{x} /\hbar \}=\exp \{
   \eta \partial/\partial y\}. 
\end{eqnarray}
 We call it the ``perpendicular shift''.
We also use the ``parallel shift'' 
$\hat{P_x}(\eta)=\exp \{i\eta \hat{y}/\hbar \}$
which shifts the wavefunction by $\eta$ 
in the diffusion space $x$ \cite{ikeda96,yamada10}.

We give a brief comment on measure of time-irreversibility.
Entropy can also be used to measure the
probabilistic feature of a wavepacket. By using entropy in place of $M(t)$
we can also characterize the sensitivity of time-irreversibility to 
the perturbation. Indeed, the time-reversal characteristics 
based on the entropy give results similar to those based on the MSD.
Indeed, as is discussed in appendix \ref{app:TREQM}, 
a characterization by entropy has a clear measurement-theoretic
significance \cite{neumann96,nicolas98,nielsen00}. 
However, entropy is quantitatively less sensitive than MSD as
the measure of the broadening of wavefunction, which is the 
significant feature of diffusion phenomenon, and we do not use it 
here.  See appendix \ref{app:TREQM} for more details.

\begin{figure}[htbp]
\begin{center}
\includegraphics[width=9cm]{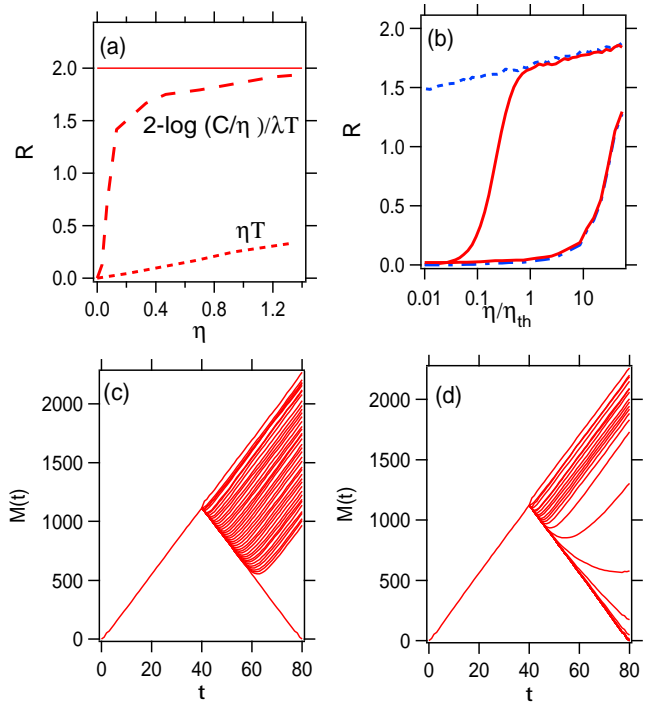}
\caption{\label{fig3}
(Color online)
(a) Schematic illustration of time-reversal characteristics $\irrev_{cl}$
as a function of the perturbation strength 
for some typical cases in classical dynamics.
See the main text for the explanation of the function form.
(b) Typical time-reversal characteristics for classical and quantum 
SM with normal diffusion obtained by time-reversal tests. 
Note that the horizontal axis is in logarithmic scale in arbitrary unit.
Typical time-reversal tests for the normal diffusion in 
(c)classical and (d)quantum SM with $K=6$, $\hbar=\frac{2\pi 121}{2^{17}}$.
The bottom curve in panel (b) denotes the typical result 
for the integrable case in SM.  (The first presentation of
time reversed dynamics like (c) and (d) is seen in 
Fig.1 of Ref.\cite{ikeda96}. See also Ref.\cite{sokolov08}.)
}
\end{center}
\end{figure}

Before going on to quantum time-reversal characteristics, let us consider 
the behavior of the time-reversal characteristics $\irrev_{cl}$
in the classical dynamics of Eq.(\ref{unitary}), which is given
by the mapping rule $(q,p)~\to~(q',p')$, namely, 
\begin{eqnarray}
\label{clsmap}
p'&=&p-V'(q+H_0'(p)/2),  \\
\nonumber q'&=& q+(H_0'(p)+H_0'(p'))/2, 
\end{eqnarray}
First, we discuss an integrable motion
where the representation by action-angle variables $(\theta, I)$
is canonical, which is transformed into $(q,p)$-space by
a canonical transformation $(q,~p)=(Q(\theta,I),~P(\theta,I))$,
where $Q$ and $P$ are $2\pi$ periodic functions of $\theta$ \cite{integrable}.
   In the integrable motion, the action is invariant and the angle exhibits
a free motion at a constant angle velocity dependent only upon 
the action: $(\theta_t,~I_t)=(\theta_0+\omega(I_0)t,I_0)$, 
where $\omega(I_0)=\frac{\partial H_0}{\partial I}|_{I=I_0}$.
At $t=T$ the applied perturbation in the $(q,p)$ plane makes the action-angle 
variables shift as $(\theta_T, I_T) \to (\theta_T+c_1\eta,I_T+c_2\eta)$, 
where $c_1,c_2$ are appropriate constants. 
Then, after time-reversed evolution for time $T$ 
 the trajectory returns to recover the initial state
\begin{eqnarray}
\theta_{2T} &=& \theta_0+c_1\eta+T(\omega(I_0)-\omega(I_0+c_2\eta)), \\
\nonumber I_{2T} &=& I_0+c_2\eta.
\end{eqnarray}
The deviation is evaluated as 
\begin{eqnarray}
|p_{2T}-p_0|\sim |\frac{\partial P(I_0,\theta_0)}{\partial \theta} 
\frac{\partial \omega'(I_0)}{\partial I}c_2 \eta T|.
\end{eqnarray}
Thus the time-reversal characteristics are 
\begin{eqnarray}
\label{clsR1}
   \irrev_{cl}=\frac{|M_\eta(2T)-M_0(2T)|}{M_0(T)} \sim \eta T.
\end{eqnarray}
The difference linearly increases with the reversal time $T$.
Therefore, we can control the accuracy of the system's return 
to the initial state
by controlling the magnitude of the perturbation strength as
\begin{eqnarray}
    \eta \sim 1/T.
\end{eqnarray}
%
See Fig.\ref{fig3}(a) for an illustration of the $\irrev_{cl}$.

On the other hand, in the case of chaotic motions, 
the perturbed orbit follows the unperturbed one only for a short period of time,
and the deviation of the former from the latter grows exponentially as 
$ d(\tau) \sim \eta \e^{\lambda \tau}$, where $\tau \equiv t-T$ for $t>T$
and  $\lambda$ is the Lyapunov exponent, up to the time 
$\tau >\tau_d$. $\tau_d$ is defined as time when $d(\tau)$ grows up to $O(1)$, 
namely $d(\tau_d) \sim O(1)(=C)$.
We refer to $\tau_d$ as the delay-time hereafter, 
because it means the time required for the loss of memory in the dynamics, beyond which 
diffusion motion is recovered in backward time-evolution at the same diffusion constant 
as the forward time-evolution due to the time reversal symmetry of the system.
  Accordingly, after the
reversal-time the increment $M_\eta(t)-M_0(T)$ increases 
like $M_\eta(t)-M_0(T) = D(\tau-\tau_d)$ for $\tau>\tau_d$.
Consequently, in the chaotic dynamics the time-reversal characteristics 
can be evaluated as 
\begin{eqnarray}
\label{clsR2}
\irrev_{cl} \sim 2-\frac{\tau_d(\eta)}{T},
\end{eqnarray}
where $\tau_d(\eta)=\frac{\log(C/\eta)}{\lambda}$. Thus 
we have to keep $\eta$ exponentially as small as
\begin{eqnarray}
\label{clsth}
   \eta \sim C \e^{-\lambda T}, 
\end{eqnarray}
if we would like to control the system to recover the 
time-reversibility.
An illustration of the time-reversal characteristics $\irrev_{cl}(\eta)$ for classical 
systems are shown in Fig.\ref{fig3}(a).

Now we return to the quantum problem and compare the quantum
result with the classical one. 
%
Figure \ref{fig3}(c) and (d) show typical time-reversal tests executed by using
classical and quantum SM in the normal diffusion region, respectively, 
and the classical time-reversal characteristics $\irrev_{cl}$ and quantum one
$\irrev$ are shown in Fig.\ref{fig3}(b).
The  quantum $\irrev$ well coincides with classical $\irrev_{cl}$
in large $\eta$ region, but the quantum $\irrev$ abruptly decreases when
$\eta$  becomes smaller than a certain threshold value, which we denote
by $\eta_{th}$.  It seems to approach algebraically to zero for $\eta \to 0$.
This result imply that in the quantum dynamics the perturbation 
does not disturb the time-reversibility of the system for the  
perturbation strength smaller than the threshold, $\eta < \eta_{th}$. 
This is in sharp contrast to classical dynamics for which control with exponentially small 
perturbation strength is required $\eta$
in order to attain the time-reversibility. Quantum systems are more
stable than the classical counterpart for $\eta < \eta_{th}$.
 In the next section, we give the qualitative evaluation of $\eta_{th}$.

\subsection{Time-reversal characteristics}
\label{subsect:irrev-chara}
In this section we examine the time-reversal test for quantum SM and
PAM in the normal diffusion region  to evaluate the time-reversal
characteristics of the systems.

\subsubsection{Case studies :standard map and perturbed Anderson map}
\label{subsubsect:SM-PAM}
We investigate time-reversal characteristics of 
diffusive quantum state by using SM, which has a classical limit, 
and PAM, which has no classical limit. As shown in Fig.\ref{fig3}, in 
quantum SM the feature of time-reversal characteristics
approaches classical one for relatively large $\eta$, if the system is
classically chaotic. Apparently there is a threshold $\eta_{th}$ 
of the perturbation strength, 
below which the system shows an intrinsic quantum characteristic.
It is expected quite naturally that the threshold $\eta_{th}$ 
is related to Planck constant $\hbar$.

In the case of SM, the wavepacket diffuses in the momentum space (i.e.,
$x=p$), and the wavepacket diffuses to cover the range of $x$ with 
width $\Delta x(T)=\sqrt{M(T)}$ at the reversal time.
So the perpendicular perturbation Eq.(\ref{pershift}) 
shifting the quantum state 
in the $y(=q)$ space by $\eta$ sweeps the phase space over the area 
$A=\eta \Delta x(T)=\eta \sqrt{M(T)}$.
Then, the shifted quantum state may be distinguished classically
from the original state if the number of quantum states contained 
in the swept area $A/h$ is larger than unity, and the separation of
orbit from the shifted state deviates classically from the original
state. 
Thereby, the threshold $\eta_{th}$ of the perturbation strength 
is estimated as, 
\begin{eqnarray}
\label{etath}
   \etath = \frac{2\pi \hbar}{\Delta x(T)}.
\end{eqnarray}
On the other hand, in the case of the parallel perturbation shifting 
the quantum state in the $x(=p)$ space by $\eta$,  the wavepacket fills 
the full domain of definition of $y(=q)$. Recall that it is defined 
by $0\leq q \leq 2\pi$.
Then the sweep area $A=2\pi\times\eta$, and the threshold perturbation
strength is given by
\begin{eqnarray}
\label{etath2}
   \etath = \hbar.
\end{eqnarray}
Hereafter, we mainly use the perpendicular $\eta-$shift as 
the perturbation, since $\eta_{th}$ depends on both $\hbar$ and $T$ 
and can be varied in two ways.
The results of the time-reversal characteristics
for the parallel $\eta-$shift are given in appendix \ref{app:parashift}.
  The above evaluation for $\etath$ can obviously be extended to general 
quantum systems. On the other hand, Eq.(\ref{etath}) can be justified by
semiclassical theory in case of SM because it has a classical limit. 
The time-reversal characteristics become classical in the condition that
the difference among the magnitudes of action integrals associated with 
classical orbits contributing to the semiclassical wavepacket is large
enough to avoid the quantum interference effect, which yields Eq.(\ref{etath}).
\begin{figure}[htbp]
\begin{center}
\includegraphics[width=9cm]{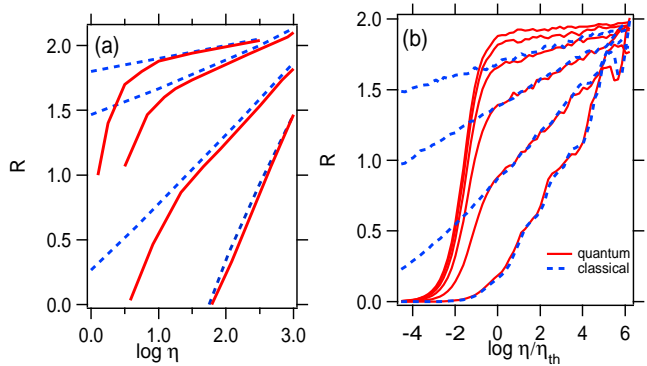}
\caption{\label{fig-SM-Reta1}
(a)Schematic illustration for the time-reversal characteristics 
of quantum and classical cases as a function of the perturbation strength $\eta$ 
at several reversal time $T$.
The axes are taken in arbitrary unit.
(b) Time-reversal characteristics of SM at several reversal 
 times $T=5,10,20,40,80$ from below.
The parameters are $K=6$, $\hbar=\frac{2\pi 121}{2^{17}}$. 
Note that the horizontal axis is scaled by LQPU. 
The time-reversal characteristics of classical cases 
are well described by Eq.(\ref{clsR2}).
}
\end{center}
\end{figure}

In Fig.\ref{fig-SM-Reta1}(a) both quantum and classical 
time-reversal characteristics of SM 
are illustrated for various reversal times $T$.
It suggests that the classical $\irrev_{cl}$ gradually increases with $T$ 
and approaches $\irrev=2$ for $T \to \infty$. 
The quantum  $\irrev$ rapidly decreases with decrease of $\eta$ in the 
{\it quantum region, $\eta < \eta_{th}$} as if it makes a hole in the 
vicinity of $\eta \sim 0$, which reflects the quantum stability against 
the perturbation.
However, as $\eta$ exceeds the threshold $\etath$ the quantum 
$\irrev$ readily reaches the classical $\irrev_{cl}$.
In particular, in the limit of 
$T \gg \tau_d(\eta_{th})$, i.e. $T \gg T_{th}\equiv -\log(C/\hbar)/\lambda$, 
$\irrev$ reaches to 2 as $\eta$ exceeds $\eta_{th}$, which means that
as $\eta>\etath$ the system completely loses the memory 
to return to the initial state and changes to the backward diffusion.  
Thus we can see the significance of the threshold $\etath$ 
as the least quantum perturbation unit (LQPU) above which the
system become entirely irreversible. 
A relation between the quantum fidelity and the  LQPU has been discussed
 by Sokolov {\it et al}\cite{sokolov08}.

It is reasonable to express the time-reversal characteristics
as a function of scaled perturbation strength $\eta/\etath$ 
to eliminate the explicit dependence of $\etath$ upon $T$ and $\hbar$.
Figure \ref{fig-SM-Reta1}(b) shows the actual quantum 
and classical $\irrev$ as a function of the scaled perturbation strength 
$\eta/\etath$ for various $T$'s in SM with $K=6$.

\begin{figure}[htbp]
\begin{center}
\includegraphics[width=8cm]{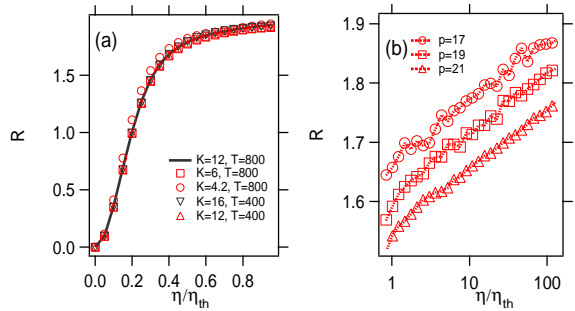}
\caption{\label{fig-SM-Reta2} 
(Color online)
Time-reversal characteristics for SM with some different parameter sets.
(a)The quantum region ($\eta < \eta_{th}$) is shown in real scale.
(b)The post quantum region ($\eta > \eta_{th}$) for $T=40$ 
is shown in logarithmic scale.
The parameters are $K=6$ and 
$\hbar=\frac{2\pi 121}{2^{p}}, p=17,19,21$
 from below. 
Hereafter we quote the data of 
$K=12,T=800$ as a reference curve.  
}
\end{center}
\end{figure}
The results clearly manifest the presence of the ``quantum hole'' for
$\eta/\etath\leq 1$. In the following, we call the region 
$\eta > \etath$ {\it post quantum region}, 
in which we can observe classical behavior if the system has the
classical limit. As is indicated by Fig.\ref{fig-SM-Reta1}(b) the $\irrev$ is 
accompanied by some irregular fluctuation in the post-quantum region,
which suggest the chaotic fluctuation inherent in the recovered
diffusive motion.

Figure \ref{fig-SM-Reta2} shows $\irrev$ for various parameter sets of SM.
In the post quantum region, $\eta > \eta_{th}$, $\irrev$ approaches the 
common line $\irrev=2$ in proportion to $1/T$ slowly. 
Also in the quantum region $\eta < \eta_{th}$, the results suggest that
$\irrev$ as a function of $\eta/\etath$ converges to a limit as $T\to \infty$. 
Moreover, it approaches a common curve if we take the limit
$\hbar \to 0$ and/or $K \to \infty$ in which an ideal quantum normal 
diffusion is realized. The characteristics of convergence 
are given in appendix \ref{app:SM-curve}.
These facts strongly suggest that in SM the time-reversal 
characteristics represented by the scaled $\eta$ becomes the same 
common curve in the limit of normal diffusive motion.

We also numerically examined the curves of $\irrev$  vs $\eta/\eta_{th}$ for SM
with various different periodic potential $V(q)$ and confirmed
that all curves coincide in the large limit of $T$ 
if the normal diffusion is maintained in the range $0<t<2T$.

In the following we use a well-converged curve of $\irrev$ vs $\eta/\etath$ 
of SM as the {\it reference curve} 
to compare with $\irrev$ of other systems. 
Thus we use $\irrev$ taken at $T=1200\gg \tau_d(\eta_{th})$ for SM with $K=12$
as the reference curve.

Figure \ref{fig-SM-Reta2}(b) shows the time-reversal 
characteristics in the post quantum region
for some cases with different Planck constant.
It is shown that in the post quantum region the slope of the plots is 
insensitive to the change of the Planck constant, this is,  
it is determined only by the Lyapunov exponent 
once the quantum dynamics is classicalized.
\begin{figure}[htbp]
\begin{center}
\includegraphics[width=9cm]{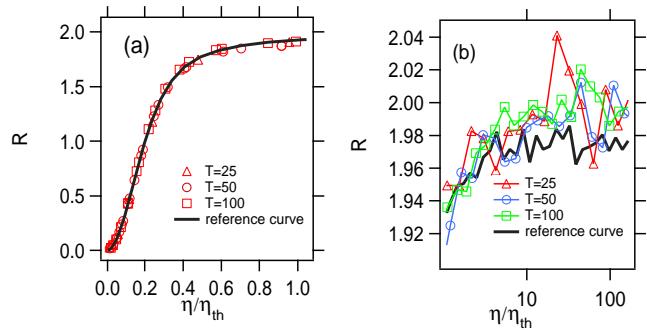}
\caption{\label{fig6} 
(Color online)
Time-reversal characteristics (as a function of 
the scaled perturbation strength) 
for PAM at several reversal times $T=25,50,100$. 
The parameters are $M=3$, $\epsilon=0.5$. 
(a) The quantum region is shown in real scale, and 
(b) the post quantum region is shown in logarithmic scale.
The reference curve is also shown.
}
\end{center}
\end{figure}

\begin{figure}[htbp]
\begin{center}
\includegraphics[width=8cm]{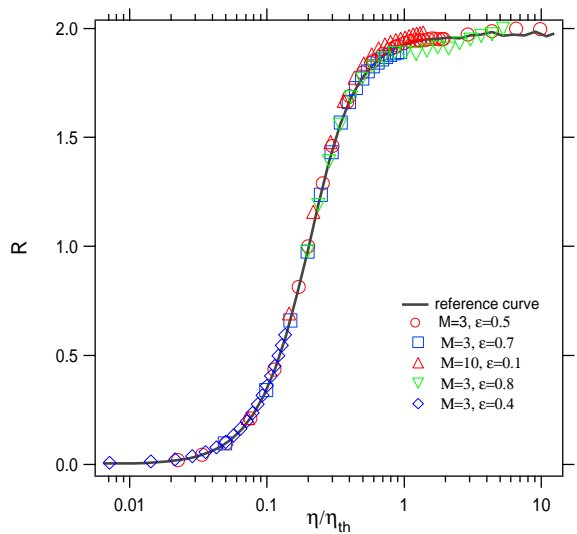}
\caption{\label{fig7} 
(Color online)
Time-reversal characteristics as a function of the scaled perturbation strength 
for PAM with some parameter sets at $T=200$.
The reference curve is also shown.
 }
\end{center}
\end{figure}

Next, we investigate the time-reversal characteristics of PAM, which has no
classical limit.  We compare it with the time-reversal characteristics
of the reference curve of SM. In PAM the normal diffusion occurs in 
the position space (namely $x=q$) quite differently from SM. 
  In the case of PAM, $\eta_{th}=\hbar$ for the parallel $\eta-$shift 
because $x=q$ and $y=p$. 
The physical origin of the normal diffusion in PAM is not chaotic
dynamics like SM but destruction of Anderson localization by harmonic 
perturbations. As seen in Fig.\ref{fig6} the time-reversal characteristics of 
PAM also have the asymptotic limit for $T \to \infty$. 
The calculated $\irrev$ for some parameter sets in PAM that generate 
normal diffusion in the $q-$space is given in Fig.\ref{fig7}. 
The plots of time-reversal characteristics coincides very well with 
the reference curve in the quantum region.
Recalling the basic difference of physical mechanisms causing the
normal diffusion in the two systems, 
such a coincidence is very surprising.
Moreover it follows that in the post quantum region the 
time-reversal characteristics 
is accompanied by erratic fluctuation as is the case in the post quantum
region of SM.

\subsubsection{Universal similarity of time-reversal dynamics in asymptotic limit
and non-universality in the way of convergence}
\label{subsubsect:universality}
\begin{figure}[!ht]
\begin{center}
\includegraphics[width=9cm]{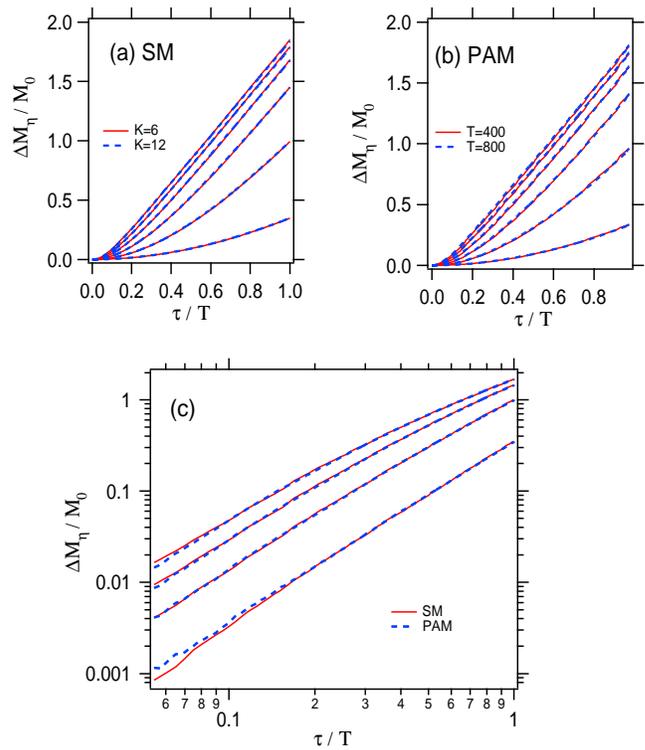}
\caption{\label{fig8}
(Color online)
Scaled separation $(M_\eta(T+\tau)-M_0(T+\tau))/M_0(T)$  
as a function of scaled time $\tau/T$ for several scaled perturbation 
strength $\eta/\eta_{th}=0.1,0.2,0.3,0.4,0.5,0.6$, respectively, from below.
(a)SM with $K=6, 12$ and $\hbar=\frac{2\pi 487}{2^{19}}, \frac{2\pi 975}{2^{20}}$, 
respectively, at $T=400$.
(b)PAM with $M=3,\epsilon=0.5$ at $T=400,800$.  
(c)Log-log plots of the data for SM and PAM at $T=400$.
}
\end{center}
\end{figure}

The fact that the $\irrev$ as the function of scaled $\eta$
asymptotically converges to a common curve means that
the time-reversed dynamics itself has an universal
scaled feature in the limit of $T \to \infty$, or more precisely $T/T_{th} \gg 1$. 
We begin with demonstrating some numerical results of 
time-reversed dynamics.
Sensitivity of time-reversed dynamics to the perturbation
is represented by the time evolution of the difference $\Delta M_\eta$ 
between the MSD's of the perturbed and the unperturbed systems
after the time-reversal operation at $T$:
\begin{eqnarray}
 \label{deltaM}
   \Delta M_\eta(T,\tau)=M_\eta(T+\tau)-M_0(T+\tau).
\end{eqnarray}
In Fig.\ref{fig8}(a) and (b), the scaled difference
$\Delta M_\eta(T,\tau)/M_0(T)$ 
is shown as a function of the scaled time $\tau/T$ for
SM at several different values of $K$ and 
for PAM at several different $T$, respectively. 
As is exemplified in Fig.\ref{fig8}(b) for PAM, the functional
form of $\Delta M_\eta(T,\tau)/M_0(T)$  as the function of $\tau/T$ 
converges to a well-defined limit in the limit $T\to \infty$.
Further, as is shown in Fig.\ref{fig8}(a) for SM, in the limit of $T\to\infty$
the scaled curves form a common shape with no regard to the parameter $K$ 
if the scaled perturbation strength $\eta/\eta_{th}$ is taken at the the
same value. Finally, as shown in Fig.\ref{fig8}(c), 
the scaled difference $\Delta M_\eta(T,\tau)/M_0(T)$ 
as a function of the scaled time $\tau/T$  is on 
the common curve independent of the system in the limit $T\to \infty$ 
if the scaled perturbation $\eta/\etath$ is taken at the same value.

This ``ultimate'' universal property of the scaled time-reversed dynamics is a
natural result of the {\it universality of the scaled time-reversality}
discussed in the previous section and {\it stationarity} of the normal diffusion 
discussed in section \ref{subsect:bayse}.
From the former, universality $\irrev$ becomes a function depending only on 
the scaled perturbation strength $\frac{\eta}{\etath(T)}$,  
 in the limit of $T \to \infty$, as
\begin{eqnarray}
\label{scale1}
     \irrev = F(\frac{\eta}{\etath(T)}), 
\end{eqnarray}
which is asymptotically independent of $T$. 
On the other hand, the stationarity of the time-reversal characteristics 
asserts that for the fixed $\eta$ the difference 
$\Delta M_\eta \equiv  M_\eta(T,T+\tau)-M_0(T,T+\tau)$ does not depend on 
the reversal time $T$ as, 
\begin{eqnarray}
\label{stationary}
     \Delta M_\eta = G(\eta,\tau),
\end{eqnarray}
where $G$ is a function depending only on $\eta$ and $\tau$.
Therefor, the difference $\Delta M_\eta$ becomes,  
\begin{eqnarray}
\label{G}
  G(\eta,\tau)=D\tau F(\frac{\eta}{\etath(\tau)}), 
\end{eqnarray}%
when we take $T=\tau$ in Eq.(\ref{G}).
Accordingly, the relation immediately follows:
\begin{eqnarray}
\label{scale2}
 \frac{M_\eta(T,T+\tau)-M_0(T,T+\tau)}{M_0(T,T)} \nonumber  \\
= \frac{\tau}{T}F(\frac{\eta}{\etath(T)}\frac{\etath(T)}{\etath(\tau)}) \nonumber  \\
= \frac{\tau}{T}F(\frac{\eta}{\etath(T)}\{\frac{\tau}{T}\}^{\chi}), 
\end{eqnarray}
where $\etath(T) \propto T^{-\chi}$. 
The index $\chi$ is determined by the type of perturbation as,
$\chi=1/2$ for perpendicular $\eta-$shift and 
$\chi=0$ for parallel $\eta-$shift. (See Eqs.(\ref{etath}) and (\ref{etath2}).)
Thus $\Delta M_\eta(T,T+\tau)/M_0(T)$ is determined only by 
the scaled perturbation strength $\eta/\etath$ and the scaled 
time $\tau/T$ for $T(>T_{th})$. 

The above result is a natural consequence of the universal
scaling of the time-reversibility and stationarity. 
The stationarity seems to mean that the past history up to the
reversal time $T$ does not influence further time evolution,
which suggests that the memory of initial state is lost, while 
the scaling property of time-reversality, which is independent 
of $T$, seems to suggests that the memory from $t=0$ to $T$ is 
maintained during the time evolved process.
Unifying the above apparently contradictory properties, we come
to the result of Eq.(\ref{scale2}) claiming that the time-reversed
dynamics have no specific time-scale.

As has been discussed, $\irrev =2$ in the post quantum region implies 
that an entire loss of memory is realized in the backward evolution and 
the backward diffusion of the wavepackets restored in the same way as the 
forward evolution without returning to the vicinity of the initial state.
In the quantum diffusive systems the spread of the wavepacket 
reaches to a macroscopic level for $T\rightarrow \infty$.
The fact that $\irrev$ reaches $2$ for $\eta>\eta_{th}$ means that
we have to control the external perturbation suffered by the wavepacket
at the strength less than the LQPU in order to make the uncertainty of 
wavefunction extended in the macroscopic level to shrink to the 
initial level. Namely, we have to control the quantum unitary 
dynamics on the {\it ultra small scale}, i.e. $\etath$
decided by the Planck constant $\hbar$. We can regard this extreme 
difficulty of recovering the time-reversibility
of a quantum system as the {\it appearance of quantum irreversibility}.

The results discussed above and in the previous section suggest that 
the time-reversal dynamics leads to a limit of convergence which is insensitive 
to the details of the system in the limit $T \to \infty$. 
However, the way of convergence with increase in $T$ depends 
on the details of the system. This is related to the short time
scale dynamics discarded in the above universality argument.

We investigate the short-time behavior of the  
time-reversed dynamics and the convergence properties 
focusing on the post quantum region. 
As discussed in Sect.\ref{subsect:irrev-test}, in the post quantum region of SM
the time-reversed dynamics is governed by classical dynamics, and 
the convergence of $\irrev$ to 2 for $\eta\gg\eta_{th}$ depends on the 
the parameter $K$ through the Lyapunov exponent, which 
controls the sensitivity of the classical chaotic dynamics.

\begin{figure}[!ht]
\begin{center}
\includegraphics[width=9cm]{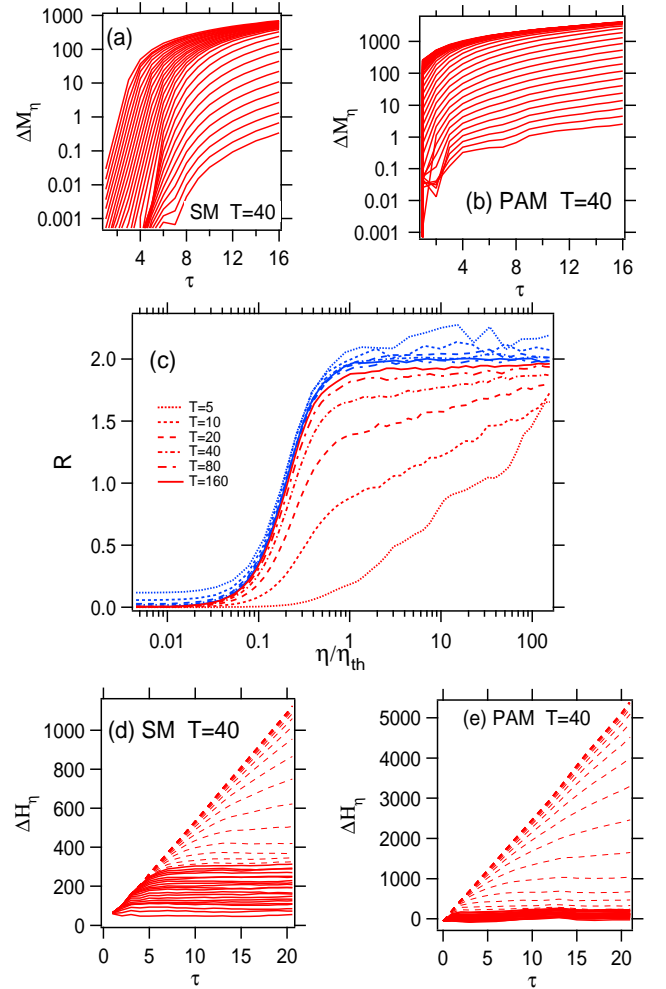}
\caption{ \label{fig9}
(Color online)
The growth of the separation $\Delta M_\eta(\tau)$ as a function of 
$\tau=t-T$ for some perturbation strength $\eta$.
(a)SM with $K=6$, $\hbar=\frac{2\pi \times 121}{2^{21}}$ and $T=40$.
(b)PAM with $M=3$, $\epsilon=0.5$ and $T=100$. 
(c)Time-reversal characteristics of quantum SM and
PAM, as functions of the scaled perturbation strength $\eta/\eta_{th}$ at several 
reversal times $T=5,10,20,40,80,160$ from below for SM and from top for
PAM, respectively.  $K=6$, $\hbar=\frac{2\pi \times 121}{2^{17}}$ 
for SM and $W=1.0, M=3,\epsilon=0.5$  for PAM.
(d) and (e) are plots of $\Delta H_{\eta}(\tau)$ 
as a function of the time $\tau$ at $T=40$, which 
are correspond to data of (a) and (b), respectively.
The dotted and bold lines represent cases 
of $\eta< \eta_{th}$ and of $\eta> \eta_{th}$,
 respectively.
}
\end{center}
\end{figure}

Figure \ref{fig9}(a) and (b) show the time dependence of $\Delta M_\eta$ for various 
perturbation strength $\eta$. The delay time $\tau_d$ can then be 
decided by $\Delta M_\eta(T,\tau)\sim O(1)$.
In case of SM (Fig.\ref{fig9}(a)), the difference $\Delta M_\eta$ shows a 
weak exponential growth with $\tau$ in the quantum region, while 
in the post quantum region it exponentially increases in the same 
way as the classical chaotic dynamics, namely 
$\Delta M_\eta(T,T+\tau)\sim \eta\e^{\lambda\tau}$. 
See appendix \ref{app:compdata-classical} for the results obeying  
purely classical dynamics. 
However, as shown in Fig.\ref{fig9}(b), in case of PAM, $\Delta M_\eta$ 
increases very rapidly in the first stage, is not exponentially but 
algebraically. In the case of SM, the delay time obeys the classical result,
 $\tau_d = \log(C/\eta)/\lambda$,  which is significantly large 
if $|\eta|\ll 1$ \cite{three-limits}.
However, in the case of PAM, the increase of $\Delta M(T,\tau)$ in $\tau$
is so rapid that $\tau_d$ is less than $1$ in the post quantum region. 
In the post quantum region of PAM the deviation of $\irrev$ 
from the converged value 2 will be much less than that of SM 
when the reversal time $T$ takes a common value.

In Fig.\ref{fig9}(c) $\irrev$ is displayed as functions 
of $\eta$ scaled by the LQPU for various $T$s.
As seen in Fig.\ref{fig-SM-Reta1}(b) 
the time-reversal characteristics of SM asymptotically 
approaches the universal curve from below when the reversal time increases.
On the other hand, in the case of PAM we can not observe a significant $T$ dependence
of the characteristics, and convergence 
to the asymptotic characteristics occurs much more rapidly than those in SM. 
As a result, in spite of the difference in the manner of convergence, 
the asymptotic limit exhibits a remarkable 
universality irrespective of the kind of the system.

We further introduce the following quantity $\Delta H_\eta$ 
in order to stress the difference of the sensitivity 
to the perturbation between SM and PAM, 
\begin{eqnarray}
 \label{deltaH}
   \Delta H_\eta(\tau)=M(T+\tau)-M_\eta(T+\tau), 
\end{eqnarray} 
where $M(T+\tau)$ denotes MSD of the forward time-evolution 
without the time-reversal operation.
In Figs. \ref{fig9}(d) and (e),
 the $\tau$ dependence of $\Delta H_\eta$ is shown for various 
perturbation strength $\eta$. 
$\Delta H_\eta$ linearly increases as $\Delta H_\eta(\tau)=2D\tau$ for  $\eta=0$. 
 It is expected that the existence of the quite different regions of
time-reversal characteristics, namely, the stable quantum 
region and unstable post quantum region, is a general and common 
feature of quantum normal diffusion irrespective of details of the model. 
  In addition, the difference between SM and PAM can clearly appear 
in the sensitivity of the $\eta-$dependence
in the post quantum region.
In the case of SM $\tau-$dependence of  $\Delta H_\eta$ becomes  flattened 
as the perturbation strength $\eta$ increases 
even in the post quantum region.
On the other hand, in the case of PAM the $\tau-$dependence of 
 $\Delta H_\eta$ drastically changes for 
the post quantum region, and it becomes insensitive to increase of $\eta$.
We also give the other plots, $2-\irrev$ v.s. $\eta/\eta_{th}$,  to express 
the difference between SM and PAM in appendix \ref{app:2-R} (Fig.\ref{afig2-R}).

\section{Normally diffusive stochastic quantum maps}
\label{sect:noise}
Unfortunately, we do not know many quantum models which exhibit the 
rigorous normal diffusion and which allow us to examine a precise numerical
simulation of normal diffusion.
To our knowledge, the coupled standard map and three-dimensional Anderson model 
are typical models which can show a normal diffusion in a rigorous sense 
\cite{abrahams10}.
Even SM and PAM, which we take as typical examples showing 
definite normal diffusion, require some conditions 
in order to realize a definite normal diffusion.
In fact, in case of SM the normal diffusion is observed only
in a finite time range beyond which the diffusion is suppressed by the localization
effect, while in the case of PAM an ideal diffusion is observed for 
sufficiently strong periodic perturbation force.

However, it is known that even quantum systems exhibiting incomplete 
diffusion can show a normal diffusion if they are perturbed by
stochastically varying random force.
(We remark that such noise-induced diffusion process is in general 
accompanied by a large fluctuation, and so an ensemble average over 
many noise processes is necessary to have reliable results.)
 In the present
section we examine the time-reversal characteristics of noise-induced
quantum normal diffusion. Here we emphasize that the term ``time-reversal''
means to reverse the whole time evolution rule including the externally
applied random force. 

Before investigating the time-reversal characteristics of 
randomly perturbed SM and AM, 
we take a quantum integrable map which shows a rigorous normal
diffusion under the perturbation of randomly varying potential, 
and examine its time reversal characteristic. 
It provides a typical example which does not have the quantum 
time irreversibility discussed so far, although the system shows 
a rigorous normal diffusion and is seemingly time-irreversible. 
Unlike SM, the quantum map has a linear kinetic energy: 
\begin{eqnarray}
\label{linear}
  H_0(p)=\omega p, ~~~~~~~V(q)=\eps \noise_t\cos q.  
\end{eqnarray}
Here $\noise_t$ is a time-dependent random variable 
with uncorrelated statistical average
 $<n_t n_{t'}>=\delta_{t,t'}$. 
Without the noise (i.e., $n_t=const.$) 
the integrable map become equivalent 
to kicked harmonic oscillator, 
which is originally proposed 
as a model of kicked charges in
a uniform magnetic field \cite{chernikov88}.
The equations of motion for the operators 
$(q,p)$ coincide with those of the classical map for the
forward-backward evolution. By setting 
$(q_t,p_t)=(U^{-t}q_0U^{t}, U^{-t}p_0U^{t})$,
it immediately follows that
\begin{eqnarray}
  q_{t+1}=q_{t}\pm\omega,~~p_{t+1}=p_{t}\pm \eps\noise_{t} \sin(q_{t}\pm\omega/2), 
\end{eqnarray}
where $+$ and $-$ are taken in forward ($t\leq T-1$) and 
backward ($t\geq T$) processes, respectively. Choosing $p_0=0$ 
as the initial state, 
\begin{eqnarray}
 p_{t}=\sum_{j=0}^{t-1}\eps\noise_{j}\sin(\omega(j+1/2)+q_0),  
\end{eqnarray}
immediately follows for $t\leq T$, which rigorously results in normal diffusion 
$M(t)=<(p_t-p_0)^2>=\eps^2 t/2$ after averaging over the random force ${n_j}$
 After the shifting 
$(q_T,p_T) \rightarrow (q_T+\eta, p_T)$, by taking the
backward process into account with the time-reversal random variable
$\noise_{j}=\noise_{2T-j-1}~~(j\geq T)$,
the time-reversal characteristics becomes 
\begin{eqnarray}
   \irrev = 4 \sin^2 \frac{\eta}{2}.
\end{eqnarray}
$\irrev$ is free from $\hbar$  and increases from the reversal state
$\irrev=0$ with $\eta$, however, it has no threshold related
to $\hbar$ above which the time-reversal characteristics $\irrev$ suddenly increase 
to order $O(1)$, which means that the system is not irreversible 
in the sense argued in the previous section. 

Our concern is whether or not the noise-induced normal diffusion may 
in general exhibit quantum irreversibility in the sense 
of the previous section.
In the following we examine the time-reversal characteristics of
stochastically perturbed SM and AM of Eq.(\ref{SSM}) type, whose control 
parameters $K$ and $\eps$ are now taken small enough
so that no spontaneous diffusion may take place.
We compare them with those in self-induced normal diffusion of SM and
PAM, which were discussed 
in the sections \ref{subsubsect:SM-PAM} and \ref{subsubsect:universality}. 
The inverse limit of localization is the ballistic motion of 
Bloch electrons in the periodic lattice. Ballistic motion
can be also converted into normal diffusion by applying
external noise. A typical example of noise-induced diffusion
in periodic lattice is seen in the Haken-Strobe model, which is described
by Eq.(\ref{kickedHaken}), namely, a one-dimensional lattice with stochastic 
potential with no spatio-temporal correlation. In addition to
the noise-driven SM and AM, we also investigate
the time-reversal characteristics of HM as a typical
example of noise-induced normal diffusion system in the periodic 
lattice.  

\begin{figure}[htbp]
\begin{center}
\includegraphics[width=9cm]{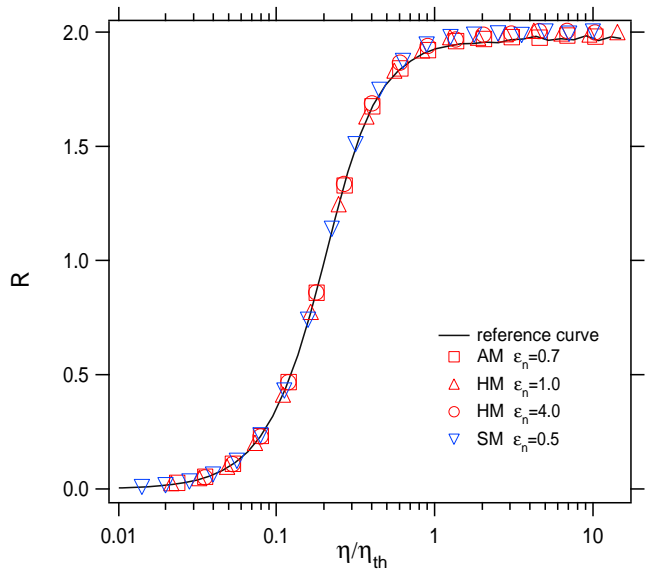}
\caption{\label{fig10} 
(Color online)
Time-reversal characteristics 
for stochastically perturbed AM with 
$\epsilon_n=0.7$, HM with $\epsilon_n=1.0,4.0$, and
SM with $\epsilon_n=0.5$ and 
$\hbar=\frac{2\pi 10001}{2^{17}}$. 
These cases all show noise driven normal diffusion.
The reference curve is also shown.
}
\end{center}
\end{figure}
It should be noted that 
in the noise-induced diffusion systems, the normal diffusion is
in general realized irrespective of the noise strength, but in
the following, we make the perturbation strength large
enough such that a well-defined normal diffusion
can be recognizable in a finite time observation. 

In Fig.\ref{fig10} we show the time-reversal characteristics of noise-induced
diffusion observed for SM, AM, and HM in strong noise region. 
See  also appendices \ref{app:strong-noise} and \ref{app:weak-noise} 
for the data of the time-reversal tests in the models.
A remarkable feature in that the time-reversal characteristics 
agrees very well with the reference curve obtained
for the {\it self-induced diffusion} in SM without the
noise source. In particular, it is surprising that the HM, 
whose diffusion mechanism does not seems to contain any dynamical 
instability like SM and AM, also exactly traces the standard 
time-irreversibility. This fact implies that even the quantum 
system which is forced to diffuse normally by the externally 
applied noise has the same time-irreversibility as the quantum 
systems exhibiting the intrinsic normal diffusion. 
In appendix \ref{app:IrrSubSuper} we compare
the time-irreversibility of the original systems such as localized
systems and ballistic systems from which the noise source is removed
with the irreversibility of the noise-induced diffusion systems 
discussed here.

In the relatively weak noise regime, the diffusion constant is small and 
the delay-time $\tau_d$ is in general extremely long, and so the converges to 
the reference curve, which is attained for $T \ll \tau_d$, is not
yet very clear. It is still an open problem whether or not the 
time-reversal characteristics of weak-noise system convergence to
the universal characteristics. 
It will be discussed elsewhere \cite{yamada11}.
 We will briefly discuss this in appendix \ref{app:noise-recover}.

\section{Summary and Discussion }
In the present paper, we investigated the time-reversal
characteristics for typical quantum maps which exhibit  
rigorous normal diffusion. Time reversality is characterized
by the difference between the forward time evolution 
and the backward time evolution which is a time-reversed process 
after an application of perturbation.

First, as examples of quantum deterministic maps, 
the standard map (SM), which is known to have a classical
counterpart, and the perturbed Anderson map (PAM), having no 
classical counterpart, are examined. 
We confirmed that there always exists the least quantum perturbation unit
(LQPU),  first introduced in Ref.\cite{ikeda96}, 
which denotes the threshold of
perturbation strength for the memory of backward evolution returning
to the initial state to be lost.  A remarkable fact is that, if the
system exhibits normal diffusion, the time-reversal 
characteristics defined as a function of the scaled perturbation
strength $\eta/\etath$ is universal in the sense that all characteristics are
plotted on the same curve in the limit of the reversal time $T \to \infty$.
In the quantum region where the perturbation strength $\eta$ is less than the 
LQPU,  the time-reversal characteristics decreases 
to zero smoothly as the scaled $\eta$ goes to zero, and makes a "quantum hole" 
around $\eta/\eta_{th}=0$. On the other hand, in the post quantum region 
where $\eta$ is larger than LQPU, i.e. $\eta/\etath>1$,  
it is accompanied by an intense fluctuation around the irreversible limit
$\irrev=2$, which means that there is a sensitive dependence on 
the perturbation strength inherent in the quantum region.
 In particular, in SM the time-reversal characteristics 
in the post quantum region 
coincided very well with those expected from the classical 
chaotic instability even for finite $T$.

The onset of normal diffusion seems to imply that the memory is lost
steadily, and the temporal {\it stationarity} is attained. On the other 
hand, we have to accept the fact that the scaled time reversality is 
universal in the sense mentioned above and the memory is conserved 
universally in the quantum region. If we accept these facts, 
we come to a conclusion that the time-reversed 
evolution process itself should obey a universal scaled dynamics 
if the perturbation strength is in the quantum region.
This conspicuous consequence was indeed confirmed numerically for 
SM and PAM. These features correspond to the most unstable class of
dynamics of the deterministic quantum systems.

We extend our time-reversal test to
the class of quantum maps which exhibit normal diffusion under the
influence of classical stochastic force. Stochastically driven 
standard map, stochastically driven Anderson map, and Haken map (HM) are
tested, and the universal time reversal characteristic examined. 
The universal time-reversed scaled dynamics confirmed for the deterministic 
quantum maps are observed also for all the stochastic quantum maps we
examined, if the stochastic force is intense enough to induce a
sufficiently rapid normal diffusion. This fact implies that even the
stochastic quantum system genuinely possesses the same degree of instability 
as the typical deterministic unstable quantum systems.  

However, whether or not the above universal time-reversal rules hold for
a weakly perturbed stochastic map, which certainly exhibits normal
diffusion but seems to share an ``integrable'' feature with the 
linear integrable map of Eq.(\ref{linear}), is still very unclear. 
There may be a threshold stochastic perturbation strength 
below which the universal time-reversal rules are
violated. This is an interesting problem.  
All the universal features mentioned above hold only in the 
converged limit $T\to \infty$, and the ways of convergence toward 
the universal time-reversible characteristics depend on the details 
of the system. 

To compare with the ideal normal diffusion, we
examined the time reversal characteristic of the localized motion,
sub-diffusive motion and finally ballistic motion.
In case of localization and subdiffusion the time-reversal 
characteristics deviates entirely from the universal curve of 
the normal diffusion, and it deviates upward from the universal curve.
On the other hand, in the ballistic motion
the time-reversal characteristics deviates downward from the universal 
curve, and the time-reversal characteristics asymptotically vanishes 
as the reversal time $T$ increases.
The behavior suggests that the ballistic state is essentially 
time-reversible as is the case in the integrable system.
An application of external stochastic force transforms the localized
and ballistic motions into a normal diffusion, and then
the time-reversal characteristics approaches close to the universal
curve from above in the localized case and 
from below in the ballistic cases, respectively.

Due to the recent development of laser techniques, 
the kicked rotor systems can now be implemented 
experimentally with ultracold atoms subjected 
to nearly resonant laser pulses.  The time-reversal 
experiment proposed in the present paper can in 
principle be tested experimentally in atomic 
optic experiments \cite{haug05}.
However, the direct observation of the time-reversal 
characteristics seems to be still rather difficult
because in order to observe ideal diffusion
in the momentum space it requires extremely long 
spatio-temporal coherence of laser pulses.

In the present paper, we focused on the time-reversibility of 
the quantum dynamics only in the time-discrete systems (quantum maps).
In a forthcoming separated paper we will compare the 
results with those of the time-continuous quantum systems  
like the Haken model and the perturbed Anderson model in which normal 
diffusion can takes places. The problems of weakly perturbed 
stochastic quantum maps mentioned above will be discussed 
there again.

\appendix

\section{A universality in parallel $\eta-$shift}
\label{app:parashift}
In this appendix, the time-reversal characteristics for the parallel
$\eta-$shift is discussed. ($\eta-$shift in $q-$space in SM, and 
$\eta-$shift in $p-$space in PAM, respectively.) 
It should be noted that in the direction of $y$-space perpendicular to 
the diffusion space $x$, the periodic boundary conditions imposed
on the system is $\Delta q=2\pi$ for SM and $\Delta p=2\pi \hbar$ for
PAM, respectively. Therefore, according to the argument given in
Eqs.(\ref{etath}) and (\ref{etath2}), the LQPU's are $\eta_{th}=\hbar$ in 
SM and $\eta_{th}=1$ in PAM, respectively. 
　
\begin{figure}[!ht]
\begin{center}
\includegraphics[width=9cm]{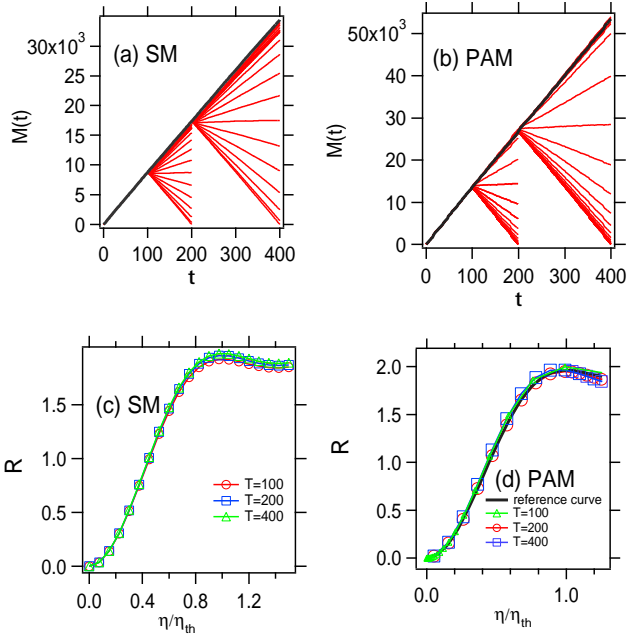}
\caption{ \label{afig3}
(Color online)
 Time-reversal experiments with the parallel $\eta-$shift for 
(a)SM with $K=12$, $\hbar=\frac{2\pi 973}{2^{20}}$ (b)PAM with $M=3$, $\epsilon=0.5$ 
at several reversal time $T$.
(c) Time-reversal characteristics for SM as a function 
of the scaled perturbation strength
at several reversal times $T=100,200,400$. 
(d)Time-reversal characteristics for PAM 
with $M=3$, $\epsilon=0.5$ at $T=100,200,400$.
The result for SM with $K=12$, $T=400$ is also shown by bold line in the panel (d) 
as a reference. Note that the horizontal axes of panel (c) and (d) are real scale.
}
\end{center}
\end{figure}
Figure \ref{afig3}(a) and (b) show the time reversal experiments 
with parallel $\eta-$shift for SM and PAM, respectively.
It is found that the backward evolution after application of the
perturbation at the reversal time shows almost linear $\tau-$dependence 
which is remarkably different from the case of perpendicular
$\eta-$shift. 

In Fig.\ref{afig3}(c) and (d), it is shown that in the quantum 
region the time reversal characteristic 
are well scaled by a single common curve for SM and
PAM.  However, in the fully post quantum region $\eta/\eta_{th}\gg 1$
the results for SM and PAM does not always coincide with each other, 
and there seems to be a significant fluctuation.

\begin{figure}[!ht]
\begin{center}
\includegraphics[width=9cm]{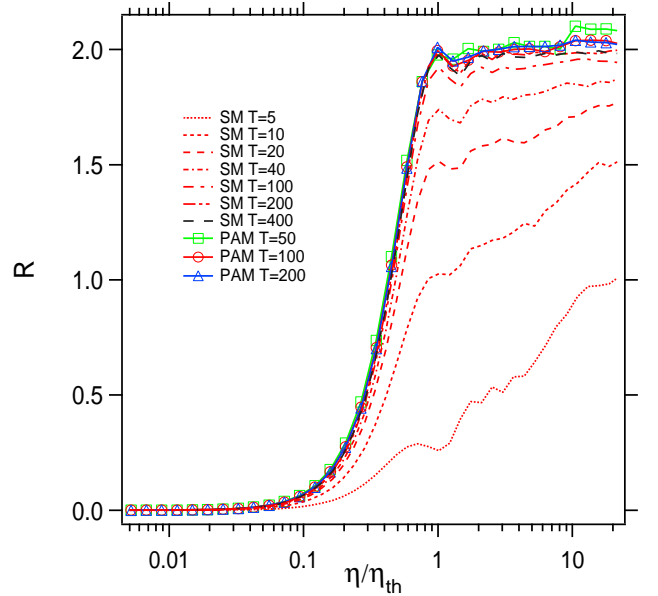}
\caption{ \label{afig13-2}
(Color online)
 Time-reversal characteristics with the parallel $\eta-$shift for 
SM and PAM as a function of scaled perturbation strength $\eta/\eta_{th}$ 
at several reversal times $T$.
$K=12$, $\hbar=\frac{2\pi 973}{2^{20}}$ for SM 
and $M=3$, $W=1$, $\epsilon=0.5$ for PAM.
The results for $T=10,20,40,80,160,400$, respectively from below for SM and 
for $T=50,100,200$ for PAM are shown.
Note that the horizontal axis is logarithmic scale.
}
\end{center}
\end{figure}

We investigate more details of the time reversal characteristic 
of the parallel $\eta-$shift. Figure \ref{afig13-2} shows the 
logarithmic plots of the time reversal characteristic
for a wide  range of $\eta/\eta_{th}$.
In the case of SM, the slow convergence due to the $1/T$ dependence is 
observed in the post quantum region as in the case of perpendicular 
$\eta-$shift. In the case of PAM the convergence is so rapid that $T$
dependence is not visible.
In addition, the presence of periodic oscillation around $\eta/\eta_{th} \sim 1$ 
is a notable feature probably due to the effect of the periodic boundary 
condition imposed for the coordinate perpendicular to the shift
coordinate, which is different from the case of the 
perpendicular $\eta-$shift. It is surprising that the oscillating
structure around the threshold
$\eta \sim \eta_{th}$ also converges to a universal $\eta/\eta_{th}$ 
dependence in the limit $T \to \infty$.

Finally we claim that the remarkable $\tau-$linear dependence of the
backward evolution depicted in Fig.\ref{afig3} is a natural consequence of 
the scaled universality (Eq.(\ref{scale2})) of the time-reversed dynamics,
 which should hold also for the parallel shift. Indeed, by setting $\chi=0$,
Eq.(\ref{scale2}) predicts the linear $\tau-$dependence of the backward 
evolution, which is just the manifestation for the universal feature
of the scaled time-reversed dynamics in the case of parallel shift.

\section{Time reversal experiment and quantum measurement} 
\label{app:TREQM}
An interpretation based on the measurement theory is possible for
the time-reversal experiments. 
We select an arbitrary eigenstate $|x_0>$ of the coordinate
$\hat{x}=\sum_x x |x><x|$, along which the diffusion takes place,   
 and observe the unitary time evolution from it as $|\Psi(T)>=U(T)|x_0>$. 
Then we consider a measurement process of measuring 
an observable $\hat{X}$ for the wavepacket $|\Psi(T)>$. 
The best measurement which gives no statistical fluctuation
is achieved if we choose the observable 
\begin{eqnarray}
\label{aphatX}
\hat{X}=U(-T)\hat{x}U(T)=\sum_{x}U(-T)|x><x|U(T), 
\end{eqnarray}
Then, at time $T$, we always measure $X=x_0$ with the probability of 100$\%$.
This forms the ``best'' measurement with the least fluctuation
which gives the least information entropy of the measurement
\begin{eqnarray}
   S=-\sum_{X}<X|\rho|X>\log<X|\rho|X>,
\end{eqnarray}
where $|X>$ is the eigenstate of $\hat{X}$ and $\rho(T)$ is
the density operator of the system, where 
$\rho(T)=U(T)|x_0><x_0|U(-T)$ in the present case. 
It is null and equals to the von Neumann entropy
\begin{eqnarray}
   S_{vN}(T)=- {\rm Tr} \rho(T) \log \rho(T),
\end{eqnarray}
which gives the lowest bound of the information entropy
of the measurement process. Thus the measurement
for Eq.(\ref{aphatX}) provides an ideal measurement process 
making the information entropy minimum. 

We are then interested in how a small perturbation $\hat{P}(\eta)$
makes the unperturbed state $U(T)|x_0>$ dirty, by observing
the perturbed state by the best measurement process for the
unperturbed state. Then the probability of measuring $X=x$ 
is $P_x(T)=|<x|U(-T)\hat{P}(\eta)U(T)|x_0>|^2$, which defines the measurement 
entropy
\begin{eqnarray}
\label{apentropy}
   S_\eta(T) =-\sum_{x} P_x(T) \log P_x(T).
\end{eqnarray}
Since $U(-T)\hat{P}(\eta)U(T)$ is nothing more than the time-reversal
process, $P_x(T)$ is the probability of finding the system 
at $x$ when the time-reversal evolution is finished at $t=2T$, 
and so $S_\eta$ is the information entropy of the time-reversed wavepacket 
which is measured by $\hat{x}$, and should be written as $S_\eta(2T)$ in
the context of the time-reversal test. Inversely speaking, 
$M_\eta(2T)=\sum_x P_x(T)(x-<x>)^2$, which we often used as 
the measure of deviation from the exactly time-reversed state,
can be interpreted as the squared average of 
the relative fluctuation at the best measurement.


\begin{figure}[!ht]
\begin{center}
\includegraphics[width=9cm]{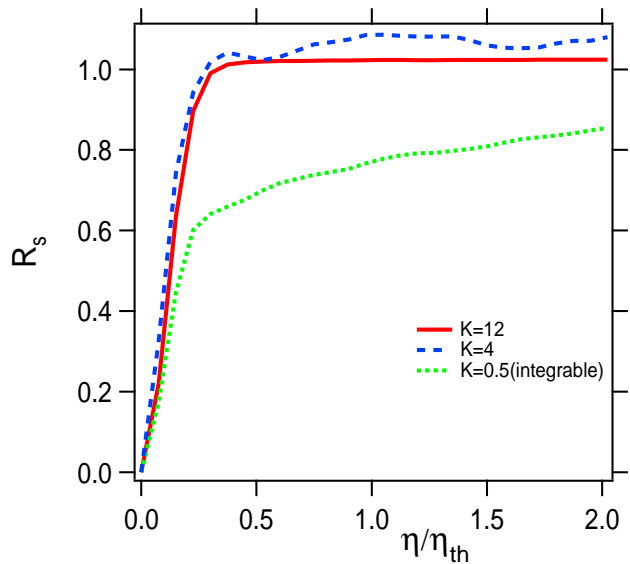}
\caption{\label{afig4}
(Color online)
The entropic time-reversal characteristics $\irrev_s$ obtained by time
evolution of entropy $S(T,\eta)$ for SM.
The results for 
the quasi-integral($K=0.5$,$T=200$), 
diffusive($K=12$,$T=100$) and  
localized($K=4$,$T=800$) cases are plotted. 
}
\end{center}
\end{figure}

The time-reversal characteristics $\irrev$ can be also defined 
by using the measurement entropy (\ref{apentropy}).
Then the entropic time-reversal characteristics $\irrev_s$ is also 
defined as 
\begin{eqnarray}
  \irrev_s(\eta,T) \equiv \frac{S(\eta,2T)-S(\eta=0,2T)}{S(\eta=0,T)}, 
\end{eqnarray}
instead of time-reversal characteristics $\irrev$ in the main text,
where the normalizing entropy $S(\eta=0,T)$ is the measurement
entropy at the reversal time for the probability $P_x=|<x|U(T)|x_0>|^2$.

Figure \ref{afig4} shows the entropic time-reversal characteristics as 
a function of the scaled perturbation strength $\eta/\eta_{th}$ for 
various quantum states in SM.  The $\irrev_s$ denotes the same tendency 
of $\irrev$.  In the quantum region $\eta < \eta_{th}$ the entropic
``quantum hole'' appears.

\section{Irreversibility of sub- and super-diffusion}
\label{app:IrrSubSuper}
In this appendix, 
we investigate the time-reversal characteristics of quantum states of
the localized, subdiffusive and ballistic time evolution and compare the results
with those of the normal diffusion.

\subsection{Time-reversal Characteristics for Localization and Ballistic motion}
\label{app:subsuper}
We employ here AM and weakly perturbed AM
as the prototypes showing the localized motion and subdiffusion, respectively.
The SM is also used as an example realizing localization if small $K$
and/or large $\hbar$ are taken.
On the other hand, we use a binary periodic system 
$v_q=1, -1, 1,-1,....$ in Eq.(\ref{AM}) as a typical example
showing the ballistic motion.
\begin{figure}[!ht]
\begin{center}
\includegraphics[width=9cm]{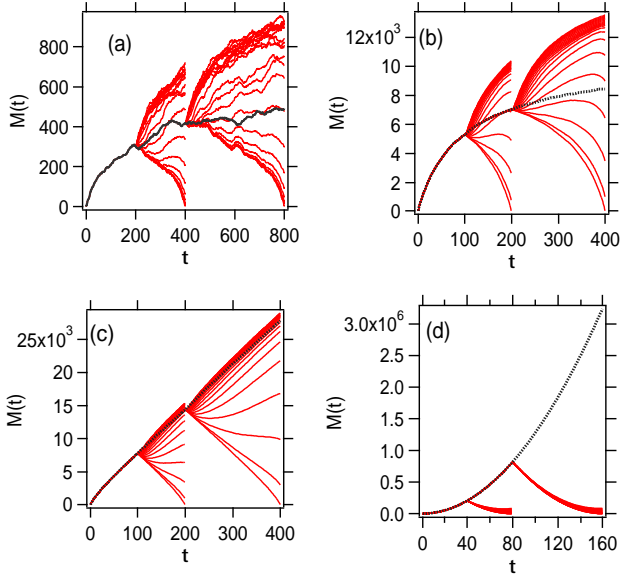}
\caption{\label{afig5}
(Color online)
Time dependence of MSD in the time-reversal experiments examined
for typical systems which do not manifest normal diffusion,  
The time-reversal backward evolutions are plotted at
two reversal times for some perturbation strength $\eta$.
(a)SM with $K=4$, $\hbar=\frac{2\pi }{2^{121}}$.
(b)AM with $\epsilon=0$.
(c)PAM with $M=3$, $\epsilon=0.05$.
(d)Case with periodic sequence for $v_n$.
}
\end{center}
\end{figure}
 Figure \ref{afig5}(a) and (b) show the time-reversal experiments 
for the localizing SM and AM. The quantum interference effect 
making the wavepacket localized is destroyed in part as the perturbation 
strength increases. As a result, for strong enough perturbation, the 
wavepacket transiently recovers the extending behavior and
the MSD increases, attaining, at the most, about 2 times 
the MSD of the localized state at
$t=2T$, namely $M_\eta(2T) \sim 2M_\eta(T)$.
Figure \ref{afig5}(c) shows the time-reversal experiments in the case 
with subdiffusive motion. 
A similar tendency to the localized states is readily seen.  
Figure \ref{afig5}(d) shows the time-reversal experiments 
examined for the ballistic evolution in a binary periodic system. 
The behavior is similar to those of the integrable systems. 

\begin{figure}[!ht]
\begin{center}
\includegraphics[width=9cm]{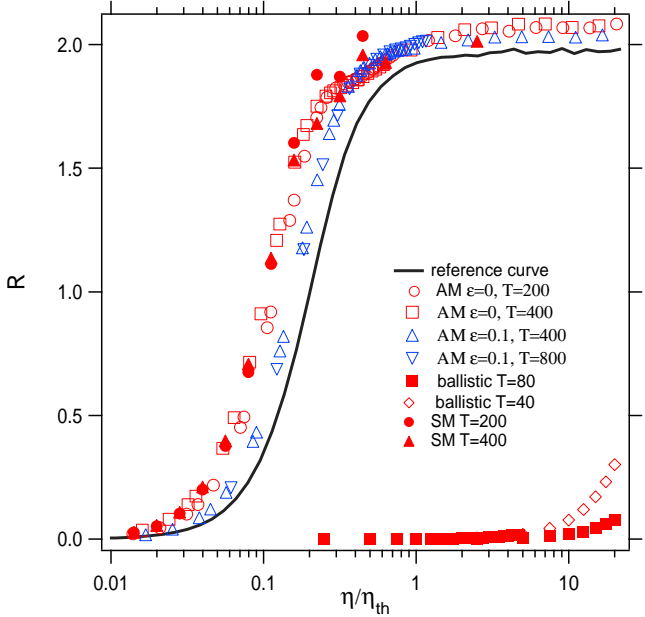}
\caption{\label{afig6}
(Color online)
Time-reversal characteristics of some cases at several reversal times.
The results for the localized (SM with $K=4$ and AM with $\epsilon=0$), 
subdiffusive (PAM with $M=3, \epsilon=0.1$)  and ballistic motion 
are plotted.
The reference curve is also shown.
}
\end{center}
\end{figure}

In Fig.\ref{afig6}, the time-reversal characteristics of the
localized, subdiffusive and ballistic motion are summarized.
In the localized state, the time-reversal characteristics 
$\irrev(\eta/\eta_{th})$ of an individual system computed at 
several reversal times seem to be on a common curve peculiar
to the individual system in the quantum region. 
However, the common curve deviates upward significantly 
from the universal characteristic curve of normal diffusion. 
This result does not mean that localized and subdiffusive states are 
more irreversible states than the normal diffusive one.

As discussed in the main text, the complete memory loss is characterized by
$\irrev=2$ in the post quantum region.
For the localized and subdiffusive states
the complete memory loss does not take place and quantum interference effect stays in 
the time evolution because the time-reversal characteristics deviates from $\irrev=2$.

On the other hand, in the ballistic motion  the time-reversal
characteristics deviates markedly downward from the universal curve,
and $\irrev$ asymptotically vanishes as the reversal time $T$ increases.
The behavior suggests that the ballistic state is essentially time-reversible
as is the case in the integrable system observed in 
Sect.\ref{subsect:irrev-chara}. 
In general, the quantum states which show the normal diffusion,
ballistic motion (and probably also superdiffusion) are all recognized
as ``extended states'' because all cases show continuous spectrum. 
However, the normal diffusion is markedly different from ballistic
motion  (and probably also from superdiffusion)
in terms of the time-reversal characteristics.

\subsection{Time-reversal characteristics of stochastically driven
  quantum maps}
\label{app:noise-recover}
As shown in Sect.\ref{sect:noise}, the localized motion
is destroyed as it is driven by stochastic noise
and a transformation into a delocalized motion
showing a normal diffusion in general occurs.
 The external stochastic noise, on the other hand, 
transforms the ballistic motion into a normal diffusion 
as is typically exemplified by the Haken map. 
Here we investigate how the time-reversal characteristic changes 
with increasing strength $\epsilon_n$ of the 
external noise. The reversal time $T$
is here taken finite, and we emphasize the strong possibility 
that all the characteristic converges to a common universal
limit even for the small $\epsilon_n$  if the limit $T\to\infty$ 
is taken first time.

The time-reversal experiments and the characteristics 
of the stochastically driven quantum maps are given in Figs.\ref{afig13}, 
\ref{afig14} and \ref{afig15}, respectively.
It can be seen that, by applying the stochastic noise, the 
localized and the ballistic motions transform into a
diffusive motion in which the time dependence of MSD increases 
in proportion to $t$, and with increase of the strength
$\eps$ and the reversal time $T$ the time-reversal characteristics 
approach to the universal curve from above in the 
localized case and from below in the ballistic case, 
respectively.

It can be expected that the stochastic perturbation introduced by 
the coupling with the external degrees of freedom can establish 
irreversibility even in the quantum systems which are not
genuinely unstable (typical example is the Ballistic map). 
Such a coupling makes the time-reversal characteristics
indistinguishable from the universal curve of deterministic 
unstable systems such as SM and PAM, which show a normal diffusion 
without stochastic perturbation.

\section{Complementary Data Sets}
\label{app:compdata}
In this appendix, we give some data which supplements the  
main text.

\subsection{Classical counterpart of the separation $\Delta M_\eta(T,\tau)$}
\label{app:compdata-classical}

\begin{figure}[!ht]
\begin{center}
\includegraphics[width=9cm]{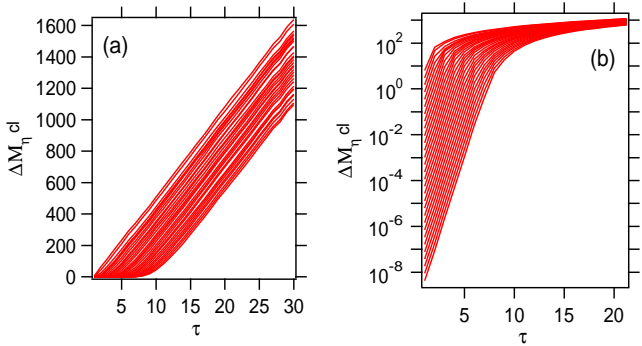}
\caption{\label{afig7}
(Color online)
Growth of the separation $\Delta M_{\eta}^{cl}(T,T+\tau)$ of classical 
SM with $K=6$ for some $\eta$'s as a function of the time $\tau(=t-T)$,
where $T=40$. 
Plots using logarithmic scale are given in (b).
The growth asymptotically approaches 
$\Delta M_{\eta}^{cl}(T,\tau) \sim 2D_{cl}(t-\tau_d)$, where $D_{cl}$ is classical
diffusion constant and $\tau_d$ is delay time, 
as the perturbation strength $\eta$ increases.
}
\end{center}
\end{figure}

In SM whose classical counterpart has chaotic phase space,
the separation exponentially increases for adequately small $\tau$ region as, 
\begin{eqnarray}
   \Delta M_\eta^{cl}(T,\tau) &=& M_\eta^{cl}(T,T+\tau)- M_0^{cl}(T,T+\tau) \\
   & \sim & \eta \e^{\lambda \tau},
\end{eqnarray}
where $\lambda$ is the Lyapunov exponent.
Here $M_\eta^{cl}(T,T+\tau)$ means MSD of classical SM at time $\tau$
after shift-perturbation at the reversal time $T$. 
Figure \ref{afig7} shows the time dependence of the actual separation 
in SM for various perturbation strengths. 
It is found that $\Delta M_\eta$ exponentially increases independent 
of the perturbation strength. 
\\

\subsection{Bayesian property for localized states}
\label{app:MSD-Bayse}
\begin{figure}[!ht]
\begin{center}
\includegraphics[width=9.5cm]{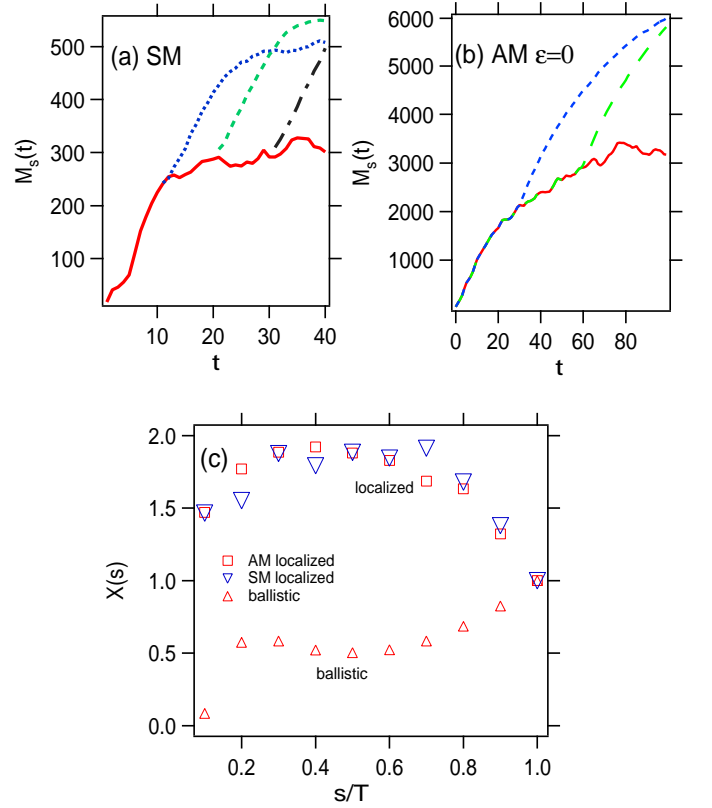}
\caption{ \label{afig9}
(Color online)
Time-dependence of MSD with observation at several intermediate
times.
$s=10,20,30$ for SM, and $s=30,60$ for PAM.
(a)SM with $K=6$, $\hbar=\frac{2\pi 305}{2^{11}}$.
(b)AM with localized behavior $\epsilon=0$. 
(c)Index $X(s)=M_s(T)/M(T)$ as a function of $s/T$ for the cases.
The curve for $s=T$ shows time-dependence of MSD without intermediate
observation as a reference. 
The result for the ballistic motion is also plotted.
In this case we take one sample for PAM.
}
\end{center}
\end{figure}

Figure \ref{afig9} shows the time-evolution of MSD 
in the localization regime, where the panels (a) and (b) are 
 AM and SM, respectively. Here, as is the case of
Fig.\ref{fig2}, a measurement of $\hat{x}$ 
is taken at several intermediate values of time $s$ taken as several 
values. Hence these are the localization version of 
Fig.\ref{fig2}

The measurement process operated at $t=s$ destroys 
the accumulated  quantum interference effect leading 
to the localization, and the evolution process is reset. 
In Fig.\ref{afig9}(c) it is shown that the index $X(s)$ 
asymptotically approaches $X(s)=1$ from above as $s \to T_f$. 
On the other hand, for the ballistic motion the index $X(s)$ 
asymptotically approaches $X(s)=1$ from below 
when $s \to T_f$, as shown in Fig.\ref{afig9}(c).  
(The MSD is not given.)

\subsection{Time-reversal test of SM}
\label{app:SM-curve}
\begin{figure}[!ht]
\begin{center}
\includegraphics[width=9cm]{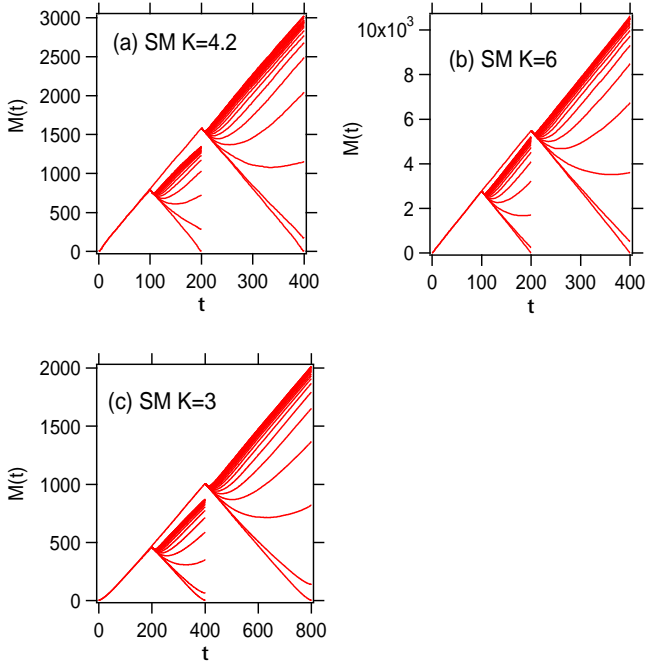}
\caption{ \label{afig10}
(Color online)
Time-reversal experiments of SM with normal diffusive behavior 
at several reversal times.
(a)$K=4.2$, (b)$K=6$ and (c)$K=3$.
The values of $\hbar$ are taken to keep the normal diffusion 
for each case, respectively.
}
\end{center}
\end{figure}

Figure \ref{afig10} shows the time-reversal experiments in SM with 
various values of the parameter sets, $\hbar$, $K$ and $T$. 
We investigate the convergence property of the time-reversal characteristics
to the universal curve. 
\\

\begin{figure}[!ht]
\begin{center}
\includegraphics[width=9cm]{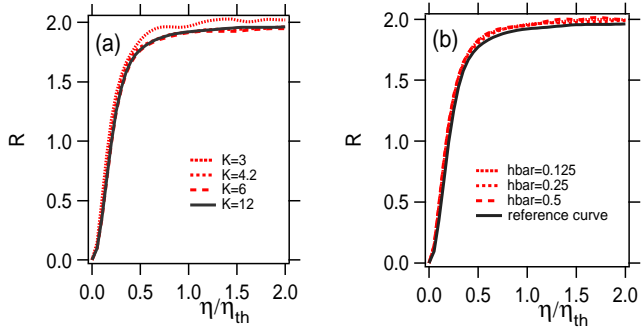}
\caption{\label{afig11}  
(Color online)
Time-reversal characteristics of SM with some parameter sets 
at reversal time $T=400$.
(a) $K=3,4.2,6,12$, $\hbar=\frac{2\pi 1947}{2^{21}}$.
(b)$K=3$, $\hbar=\frac{2\pi 61}{2^{16}}, \frac{2\pi 61}{2^{17}}, 
\frac{2\pi 61}{2^{18}}, \frac{2\pi 61}{2^{19}}$.
}
\end{center}
\end{figure}
Figure \ref{afig11}(a) shows the time-reversal characteristics 
of SM at various values of $K$
and a fixed $\hbar$, and (b) at various $\hbar$ and fixed $K=3$.
$K=3$ data do not still converge to the universal
curve because of the small value of Lyapunov exponent.

\subsection{Time-reversal test for PAM}
\label{app:PAM-curve}
\begin{figure}[!ht]
\begin{center}
\includegraphics[width=8cm]{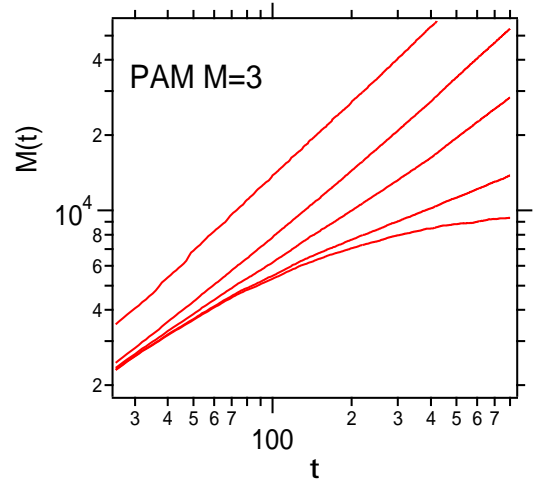}
\caption{\label{afig8}
(Color online)
Log-log plots of time-dependence of MSD for PAM with different values of 
perturbation strength $\epsilon=0,0.02,0.05,0.1,0.5$, respectively from below.
}
\end{center}
\end{figure}

Most readers will be, however, unfamiliar with the perturbed
Anderson map (PAM), so we first depict how the normal 
diffusion is attained with increase in the perturbation 
strength $\epsilon$.  As shown in Fig.\ref{afig8},
in the Anderson map (AM) the increase of MSD soon 
saturates without the perturbation and a
typical localization behavior takes place,
but the nature of time evolution changes as localization 
$\rar$ subdiffusion $\rar$ normal diffusion with increase 
in the strength $\epsilon$. We here consider the
normal diffusion regime.

\begin{figure}[!ht]
\begin{center}
\includegraphics[width=9cm]{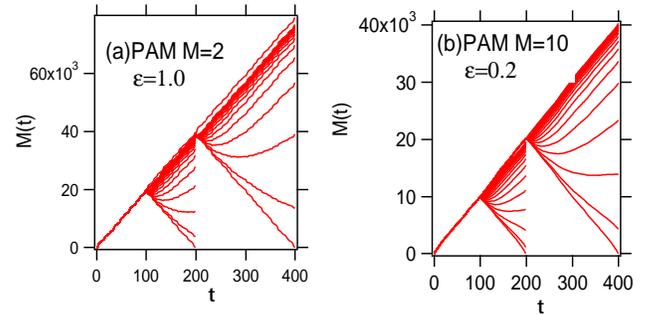}
\caption{\label{afig12}
(Color online)
Time-reversal experiments of PAM with normal diffusive behavior 
at several reversal time.
(a)$M=2$, $\epsilon=1.0$.
(b)$M=10$, $\epsilon=0.2$.
}
\end{center}
\end{figure}

Figure \ref{afig12} shows the time dependence 
of MSD for several values of the parameter sets of
PAM exhibiting a normal diffusion.
The time-reversal characteristics for the 
normal diffusion shows the universal feature
in the quantum region
independent of the diffusion coefficient, 
as shown in Fig.\ref{fig7} in the main text.

%

\subsection{$2-\irrev$ in SM and PAM}
\label{app:2-R}
\begin{figure}[!ht]
\begin{center}
\includegraphics[width=9cm]{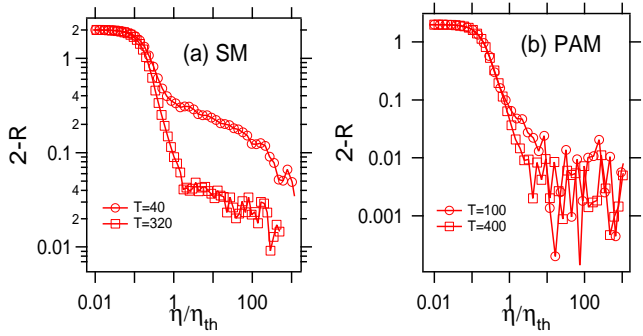}
\caption{\label{afig2-R}
(Color online)
Plots of $2-\irrev$ as a function of the scaled perturbation strength 
$\eta/\eta_{th}$ in double logarithmic scale.
(a)SM with $K=6$, $\hbar=\frac{2\pi 121}{2^{21}}$ and $T=320,640$.
(b)PAM with $M=3,2$, $\epsilon=0.5$ and $T=400$.
}
\end{center}
\end{figure}

Supposing that Eq.(\ref{clsR2}) can be used 
in the post quantum region in general,
the deviation of $\irrev$ from converged value 2 in the post quantum
region of PAM will be much less than that of SM, 
as is depicted in Fig.\ref{afig2-R}(a)
and (b), if $T$ takes a common value. Thus the sensitivity of 
the time-reversed dynamics to the perturbation is responsible 
for the convergence of the time-reversal characteristics.

We finally remark that the time-reversal characteristics plotted in
Fig.\ref{afig2-R}(a) and (b) manifests the presence 
of intense fluctuation inherent 
in the post quantum region, which contrasts with the quantum region 
in which  $2-\irrev$ decreases smoothly as the function of $\eta$.

\subsection{Strongly noise-induced normal diffusion}
\label{app:strong-noise}
\begin{figure}[!ht]
\begin{center}
\includegraphics[width=9cm]{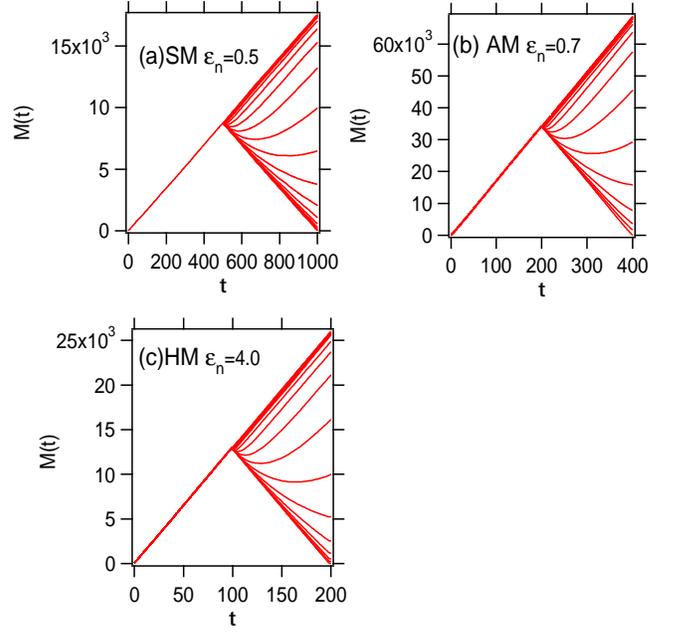}
\caption{ \label{afig13}
(Color online)
Time-reversal experiments of some 
stochastic models with noise-driven 
normal diffusive behavior.
(a) Noise-driven SM with $K=6$, $\epsilon_n=0.025$.
(b) Noise-driven AM with $\epsilon_n=0.7$.
(c) Haken map with $\epsilon_n=4.0$.
}
\end{center}
\end{figure}

Figure \ref{afig13} shows the time dependence 
of MSD for normal diffusion with large diffusion 
coefficient in noise-induced SM, PAM and HM, 
which is driven by stochastic forces with 
large strength $\epsilon_n$.
%
%
The normal diffusion 
by the backward evolution asymptotically 
approaches to the normal diffusion by the 
forward evolution, and the time-reversal 
characteristics approaches to the universal 
one as the perturbation strength $\eta$ increases.

\subsection{Weakly noise-induced normal diffusion}
\label{app:weak-noise}
\begin{figure}[!ht]
\begin{center}
\includegraphics[width=9cm]{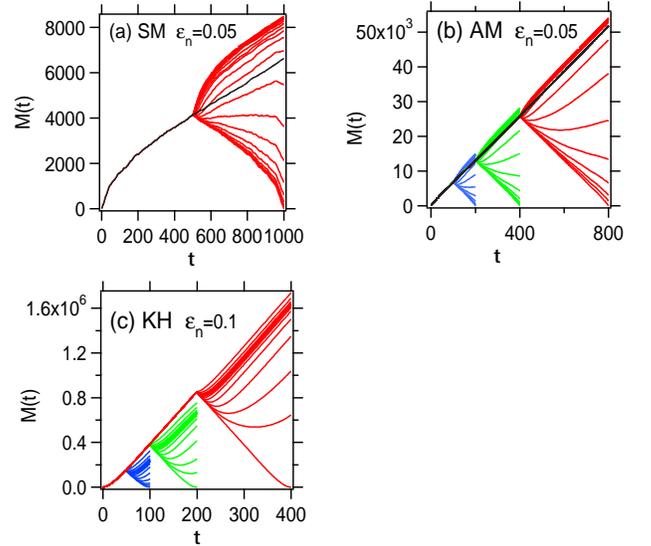}
\caption{\label{afig14}
(Color online)
Time-reversal experiments of some stochastic models without 
normal diffusive behavior due to weak noisy perturbation.
(a) Noise-driven SM with $K=6$, $\hbar=\frac{2\pi 10001}{2^{17}}$, 
$\epsilon_n=0.05$.
(b) Noise-driven AM with $\epsilon_n=0.05$.
(c) Haken map with $\epsilon_n=0.1$. 
The thick lines in the panels (a) and (b) denote the result for 
forward evolution as a reference. 
}
\end{center}
\end{figure}

In this subsection, the data of the time-reversal experiments are given 
for normal diffusion realized by the weakly noise driven SM, PAM and HM.
Figure \ref{afig14} shows the time dependence of MSD. 
Note that in weakly noise driven cases the time dependence of MSD deviates
from the typical $t$-linear dependence of the normal diffusion, 
at least in the the early stage of the time-evolution, 
although it shows the exact normal diffusion with
the $t$-linear dependence on a longer time-scale.  
However, whether the time-reversal characteristics
of the weakly noise driven system approaches to the
universal curve is still an open question.

\begin{figure}[!ht]
\begin{center}
\includegraphics[width=9cm]{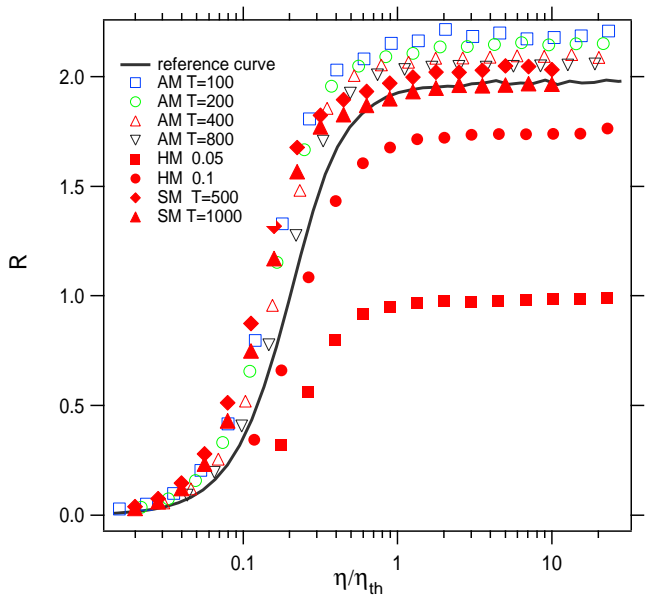}
\caption{\label{afig15}
(Color online)
Time-reversal characteristics of some stochastic models in 
pre-normal diffusive behavior due to weak stochastic perturbation.
Results for SM with $K=6$, $\hbar=\frac{2\pi 10001}{2^{17}}$, $\epsilon_n=0.05$,
 AM with $\epsilon_n=0.05$ and HM with $\epsilon_n=0.05, 0.1$ at 
 several reversal-times are plotted.
The reference curve is also shown.
}
\end{center}
\end{figure}

Figure \ref{afig15} shows the actual time-reversal characteristics 
for the weakly noise driven maps. In the Haken map the time-reversal
characteristics deviates downward from the reference curve 
due to the ballistic-like growth in the early stage, and it 
asymptotically approaches the reference curve. 
On the other hand, in noise driven SM and AM  the time-reversal
characteristics deviates upward from the universal curve.
The behavior is  similar to that in the localized state.

\section*{Acknowledgments}
This work is partly supported by Japanese people's tax via MEXT,
and the authors would like to acknowledge them.
They are also very grateful to Dr. T.Tsuji and  Koike memorial
house for using the facilities during this study.



\begin{references}

\bibitem{prigogine81}
I. Prigogine,
{\it From Being to Becoming: Time and Complexity in the Physical Sciences}
(W H Freeman, San Francisco, 1981). 

\bibitem{ikeda93}
K. Ikeda, Ann. Phys. (N.Y.) {\bf 227}, 1(1993).

\bibitem{yamada02} H. Yamada and K.S. Ikeda, 
Phys. Rev. E {\bf 65}, 046211-1-17(2002).

\bibitem{yamada99}H.Yamada and K.S. Ikeda, 
Phys. Rev. E {\bf 59}, 5214-5230(1999).

\bibitem{caldeira83}
A.O. Caldeira and A.J. Leggett,
Physica A{\bf 121},587(1983), 
 Phys.Rev.A {\bf 31}, 1059-1066(1985).


\bibitem{dittrich90} T. Dittrich and R. Graham,  
Ann. Phys. (NY) {\bf 200}, 363(1990). 

\bibitem{adachi88}S. Adachi, M. Toda and K. Ikeda,
Phys. Rev. Lett. {\bf 61}, 655-658(1988); {\it ibid}, 659-661(1988). 


\bibitem{dunlap89} 
D. H. Dunlap, K. Kundu, and P. Phillips, 
Phys.Rev. B{\bf 40}, 10999(1989). 

\bibitem{dunlap90} 
D.H. Dunlap, H.L. Wu and T. Phillips, 
Phys.Rev. Lett. {\bf 65}, 88-91(1990). 

\bibitem{huang01} 
X.Q. Huang, R.W. Peng, F. Qiu, S.S. Jiang and A. Hu,
Eur. Phys. J. B {\bf 23}, 275-281 (2001).

\bibitem{izrailev00}F.M. Izrailev, T. Kottos, A. Politi and G.P. Tsironis, 
Phys. Rev. E {\bf 55}, 4951-4963(1997); 
A. Politi, S. Ruffo and L. Tessieri, Euro. Phys. J. B {\bf 14}, 673-679(2000).

\bibitem{kawarabayashi96} 
T. Kawarabayashi and T. Ohtsuki, 
Phys.Rev. B{\bf 53}, 6975-6978(1996). 


\bibitem{yamada10} H.S. Yamada and K.S. Ikeda,
 Phys. Rev. E {\bf 82}, 060102(R)(2010).



\bibitem{haken73} H. Haken and G. Strobl, 
Z. Phys. {\bf 262}, 135(1973). 

\bibitem{capek85}V. Capek,
Z.Phys. B {\bf 60}, 101-105(1985).

\bibitem{fidelity-classical}
The fidelity defined on the basis of the phase-space
distribution function has formally quantum-classical 
correspondence. However,  it is much less convenient 
for measuring time reversibility than the mean square
displacement used in the present paper.




\bibitem{peres84}A. Peres, 
Phys. Rev. A {\bf 30}, 1610-1615 (1984).

\bibitem{benenti02} G. Benenti and G. Casati, Phys. Rev. E {\bf 65}, 066205(2002).


\bibitem{benenti03}J. Vanicek, and E. J. Heller,　
Phys. Rev. E {\bf 68}, 056208 (2003). 



\bibitem{hiller04} 
M. Hiller, T. Kottos, D. Cohen and T. Geisel,
Phys. Rev. Lett.  {\bf 92}, 010402(2004).




\bibitem{mintert05} F. Mintert, A. R. R. Carvalho, M. Kus, and A. Buchleitner,
Phys. Rep. {\bf 415}, 207(2005).


\bibitem{gorin06}
T. Gorin, T. Prosen, T. H. Seligman and M. Znidaric,
Phys. Rep. {\bf 435}, 33-156 (2006).

\bibitem{jacquod09}
Ph. Jacquod and C. Petitjean, 
Adv. Phys. {\bf 58}, 67 (2009). 



\bibitem{haug05}
F. Haug, M. Bienert, and W. P. Schleich, 
T. H. Seligman and M. G. Raizen,  
Phys. Rev. A {\bf 71}, 043803 (2005).


\bibitem{fidelity} 
Random potential/disorder type perturbations
are frequently used in time reversal experiments to investigate 
the fidelity of evolution of wavepackets, in which experiments 
the evolution is affected by the perturbation 
for the entire time-reversal period, i.e., for times from $T$
to $2T$. 


\bibitem{ikeda96}
K. Ikeda, 
{\it Time irreversibility of classically chaotic quantum dynamics}, 
145-153,  Ed. by G.Casati and B.V.Chirikov (Cambridge Univ. Press, 1996).



\bibitem{chirikov88}B. V. Chirikov, F. M. Izrailev, and D. L. Shepelyansky, 
Sov. Sci. Rev. C {\bf 2}, 209(1981); Physica D {\bf 33}, 77(1988); 
G. Casati, B.V. Chirikov, I.Guarneri, and D.L. Shepelyansky,  
Phys. Rev. Lett. {\bf 56}, 2437(1986). 


\bibitem{shiokawa95}
K. Shiokawa and B. L. Hu,  
Phys. Rev. {\bf E} 52, 2497-2509 (1995). 


\bibitem{sokolov08} 
V.V. Sokolov, O.V. Zhirov, G. Benenti, and G. Casati, 
Phys. Rev. E {\bf 78}, 046212(2008);
G. Benenti and G. Casati,
Phys. Rev. E {\bf 79}, 025201(R)(2009).


\bibitem{yamada11} H.Yamada and K.S. Ikeda,
in preparation.





\bibitem{yamada04}H. Yamada and K.S. Ikeda, 
Phys. Lett. A {\bf 328}, 170-176(2004).

\bibitem{arnold98} 
L. Arnold. {\it Random Dynamical Systems },
(Springer-Verlag, New York, 1998).





\bibitem{neumann96} J. Von Neumann, 
{\it Mathematical Foundations of Quantum Mechanics}, 
 (Princeton Landmarks in Mathematics and Physics, 
Princeton Univ Pr, 1996).

\bibitem{nicolas98} N.J. Cerfa,  and C. Adamib,
Information theory of quantum entanglement and measurement,
Physica D{\bf 120},  62-81(1998). 

\bibitem{nielsen00} 
Nielsen, Michael A. and Isaac L. Chuang, 
{\it Quantum Computation and Quantum Information}
 (Cambridge University Press, 2000). 






\bibitem{integrable}
A very few number of completely integrable maps are known. However, the situations 
discussed here is almost correct around KAM invariant tori of nonintegrable 
maps.


\bibitem{three-limits} 
There exist three parameters $T$, $\hbar$ and $\eta$.
We are considering $T$ at which the system completely mimic classical
chaotic diffusion. In other words, we are considering an ideally
diffusive system. The threshold $\eta_{th}$ is a universal parameter
observed for such system. If we take first $\eta \to 0$ and next
$\hbar \to 0$ then the system is completely reversible,
whereas in the limit $\hbar \to 0$, the system follows completely
the classical unstable dynamics.



\bibitem{abrahams10} E. Abrahams, 
{\it 50 Years of Anderson Localization} (World Scientific Pub Co Inc 2010). 


\bibitem{chernikov88}
A. A. Chernikov, R. Z. Sagdeev and G. M. Zaslavsky, 
Physica {\bf D 33}, 65 (1988).









\end{references}
\end{document}